\documentclass[journal]{IEEEtran}
\UseRawInputEncoding
\usepackage{multirow}
\usepackage{diagbox}
\usepackage{amssymb}
\usepackage{amsmath}
\usepackage{graphicx}
\usepackage{cite}
\usepackage{citesort}
\usepackage{subfigure}
\usepackage{graphicx,epstopdf}
\usepackage{epsfig}	
\usepackage{cite,graphicx,amsmath,amssymb}
\usepackage{comment}
\usepackage{algorithm}
\usepackage{algorithmic}
\usepackage{graphicx}
\usepackage{makecell}

\usepackage{amssymb}
\usepackage{amsmath}
\usepackage{cite}
\usepackage{url}
\usepackage{xcolor}
\usepackage{cite,graphicx,amsmath,amssymb}
\usepackage{subfigure}
\usepackage{citesort}
\usepackage{fancyhdr}
\usepackage{mdwmath}
\usepackage{mdwtab}
\usepackage{caption}
\usepackage{amsthm}
\usepackage{lipsum}
\usepackage{cite}

\newtheorem{theorem}{Theorem}

\newtheorem{lemma}{Lemma}

\newtheorem{corollary}{Corollary}

\newtheorem{remark}{Remark}  
\newtheorem{proposition}{Proposition}

\makeatletter
\def\ScaleIfNeeded{%
\ifdim\Gin@nat@width>\linewidth \linewidth \else \Gin@nat@width
\fi } \makeatother

\begin{document}
\title{\Huge{Relay Satellite Assisted LEO Constellation NOMA Communication System}}

\author{ Xuyang~Zhang,~Xinwei~Yue,~\IEEEmembership{Senior Member~IEEE}, Zhihao Han, Tian Li,~\IEEEmembership{Member,~IEEE}, Xia Shen, Yafei Wang, Rongke Liu,~\IEEEmembership{Senior Member IEEE}

\thanks{X. Zhang, X. Yue and Y. Wang are with the Key Laboratory of Information and Communication Systems, Ministry of Information Industry and also with the Key Laboratory of Modern Measurement $\&$ Control Technology, Ministry of Education, Beijing Information Science and Technology University, Beijing 100101, China (email: \{xuyang.zhang, xinwei.yue and wangyafei\}@bistu.edu.cn).}
\thanks{T. Li is with the 54th Research Institute of China Electronics Technology Group Corporation, Shijiazhuang Hebei 050081, China. (email: t.li@ieee.org).}
\thanks{X. Shen is with the China Academy of Information and Communications Technology (CAICT), Beijing 100191, China (email: shenxia@caict.ac.cn).}
\thanks{Z. Han and R. Liu are with the School of Electronic and Information Engineering, Beihang University, Beijing 100191, China. R. Liu is also the Shenzhen Beihang Emerging Industry Technology Research Institute, Shenzhen 518063, China (email: \{hzh$\_$95, rongke$\_$liu\}@buaa.edu.cn).}
 }




\maketitle

\begin{abstract}
This paper proposes a relay satellite assisted low earth orbit (LEO) constellation  non-orthogonal multiple access combined beamforming (R-NOMA-BF) communication system, where multiple antenna LEO satellites deliver information to ground non-orthogonal users. To measure the service quality, we formulate a resource allocation problem to minimize the second-order difference between the achievable capacity and user request traffic. Based on the above problem, joint optimization for LEO satellite-cell assignment factor, NOMA power and BF vector is taken into account. The optimization variables are analyzed with respect to feasibility and non-convexity. Additionally, we provide a pair of effective algorithms, i.e., doppler shift LEO satellite-cell assisted monotonic programming of NOMA with BF vector (D-mNOMA-BF) and ant colony pathfinding based NOMA exponential cone programming with BF vector (A-eNOMA-BF). Two compromise algorithms regarding the above are also presented. Numerical results show that: 1) D-mNOMA-BF and A-eNOMA-BF algorithms are superior to that of orthogonal multiple access based BF (OMA-BF) and polarization multiplexing schemes; 2) With the increasing number of antennas and single satellite power, R-NOMA-BF system is able to expand users satisfaction; and 3) By comparing various imperfect successive interference cancellation, the performance of A-mNOMA-BF algorithm exceeds D-mNOMA-BF.
\end{abstract}
\section{Introduction}
With the increase in the types and numbers of communicating spacecraft, the tasks for tracking and controlling mobile equipments are becoming more prominent. The amount of satellite data transmission was also increasing, which was difficult to be performed by ground measurement and control stations alone\cite{WW2021}. For the urgent demand with wide area connectivity and global access, the terrestrial cellular and satellite communication networks  faced great challenges to continue their independent development\cite{Dar2022}. The integrated air-space-ground communication systems were the development trend of future communication networks to achieve efficient resource scheduling\cite{Cui2021}. Satellite constellations utilized synergistic capabilities for global communications, navigation, environmental monitoring and other missions\cite{Ji2022}. Low earth orbit (LEO) satellite constellation  enabled any location on earth to be covered by satellites at any moment\cite{LiuS2020}. Moreover, relay satellites was able to provide measurement data and control services for satellites with other orbital altitudes\cite{Cao2018}. To obtain more warning time in response to major natural disasters\cite{Lindsey}, the relay satellites have greatly improved the efficiency of using various types as satellites.

Up to now, non-orthogonal multiple access (NOMA) technology has been applied in a variety of star-earth domains. The NOMA has shown a stronger resource allocation capability compared to orthogonal multiple access (OMA)\cite{zhang2019,Jia2019}. LEO constellations applying NOMA is able to the lower latency of multi-tasking services types in fifth-generation-advance  networks \cite{Hu2023}. Based on the above study, this paper further surveyed that the different requirements of terrestrial users can be satisfied by NOMA based LEO constellation. The authors of \cite{Jiao2020} investigated  the information delivery rate maximization limited by data queuing and power allocation in the satellite-based Internet of Things. In addition, the authors of \cite{Ge2021} designed the bandwidth compression of satellite terrestrial NOMA networks to guarantee users' fairness. Both user fairness and rate maximization are difficult to realize user traffic fitting from different directions, which encourages the optimal satellite resource allocation in various beams via NOMA.  Multiple-beam NOMA with different architectures were discussed in \cite{Wang2022A,wang2021A}, where the ground users were  imperfectly orthogonal in the more distant space. The authors of \cite{Gao2021} considered NOMA assisted multi-antenna satellite systems with imperfect successive interference cancelation (ipSIC), where the NOMA scheme was verified to obtain higher system rates. In \cite{Celik}, the ipSIC scheme suffered from the nasty effects of inter-cell interference and similar user channel gains. In practice, there will be interleaving among the satellite beams, and it is intuitive to encourage multiple users with different conical areas  performing the same NOMA codebook.

Compared to traditional terrestrial mobile networks, cooperative satellite communications have provided multi-access connectivity and flexible mobility\cite{Deng2021R,Leyva2020}. The theoretical capacity requirement was proved in \cite{Deng2021R} by the optimization algorithm of three-dimensional LEO satellites.  As a further advance, the authors of \cite{Leyva2020} integrated the LEO satellite into the fifth generation system, where the dynamic satellite link regime was more effective to reduce the transmission delay.  Multi-satellite assisted terrestrial networks were discussed in \cite{Chae2018}, which effectively improves the overall system throughput. However, the user requirement and satellite orbit information were unstable and limited in  \cite{Zhang2020Yuan}. Due to the complex atmospheric conditions, the authors of \cite{zhou2022Di} showed that the deployment location of ground stations severely affect the resource utilization of LEO satellites. In \cite{Wang2023feng}, the relay satellite brilliantly planned the trajectory of a large-scale LEO satellites and increased the transmission rate through a time extension scheme. In addition, the authors of \cite{Hills2022} discussed  spectral coexistence interference for the geostationary earth orbit (GEO) and Ku-band LEO satellite communications. In \cite{Khan2022}, LEO and GEO satellites coexisting communication scenarios were considered, where the poor high temperature effects of GEO satellite users hardly were ignored. For illustration purposes, the sharp decline of user satisfaction caused by GEO satellite gateway interference was verified in \cite{Lin2023}. As a result, GEO satellites should provide more reasonable auxiliary functions for large-scale LEO satellite constellation networks. The relay satellite was able to cover LEO satellites on a large scale and to carry out unified planning based on measurement information \cite{Miridakis2015}, which is capable of avoiding the waste of satellite orbit resources. High-orbit relay satellite enhanced the system capacity of LEO satellite communications, where the grand users were virtually free from interference \cite{Fan2016}.

To flexibly accomplish beam alignment for LEO satellite \cite{Xv2023}, a practical user-accurate positioning scheme was proposed by exploiting the internal beamforming (BF) design and external beam scheduling. Unlike high orbit satellite communications, large-scale LEO satellites more easily solved complex user-base station association matrices with BF \cite{Zhou2023}, where the shaped beams were designed arbitrarily. Moreover, the authors of \cite{Lu2019W,Wang2021QZ} proposed a robust BF scheme to overcome the interference generated for multi-beam LEO satellite networks. To serve many users within limited beams, the LEO satellites based packet BF technique made the beam center more concentrated in the user's area \cite{Wu2022D}. From the perspective of conserving satellite resources, the authors of \cite{Lin2023zhi} studied the power constrained LEO satellites communications by iterating the weight vectors of the beams.  In general, LEO satellites with beam-hopping techniques were employed serve users of different specific regions in discrete time slots \cite{zhang2023}, whereas BF enables the continuous service of users at arbitrary locations. The successively different directional beams still satisfy different types of terminals and mitigate inter-user interference.
\subsection{Motivations and Contributions}
The aforementioned investigation results have provided the basis of superior analysis on satellite networks with NOMA and BF. However, the NOMA  conveniently serves multiple mission types of user access to meet complex traffic demands in LEO constellation communications. In parallel, BF formed by the number of existing LEO satellite antennas is not sufficient for full orthogonality among users. Hence, this paper investigates whether the combination of NOMA and BF further improves the satisfaction of users. Since ground stations are limited by geographical factors, LEO constellations lack the flexibility and comprehensiveness to obtain orbital and users' information, etc. Legacy GEO satellite produces the effects of undesirable high temperatures and interference for LEO satellite communications. As a consequence, relay satellite is considered to track and measure the status information of LEO satellites anytime, and plays an important role in resolving the Doppler shift and path coordination. To the best of our knowledge, there are no related works to consider the resource optimization of the relay satellite assisted LEO constellation  NOMA combined BF (R-NOMA-BF) system, which motivates us to elaborate it.  The thesis iterates and transforms non-convexity in the optimization problem, and proposes two associated joint optimization algorithms. In general, resource optimization of R-NOMA-BF system usually leads to complex joint optimization problems. In general, resource optimization of R-NOMA-BF system usually leads to complex joint optimization problems. The optimization may not be achievable for large-scale instances owing to unaffordable complexity and time. For difficult solving mixed integer non-convex programming (MINCP) problems, the optimal solution is probably unknown. Consequently, important motivations for exploiting the algorithms are: 1) Determining the degree of difficulty of the resource optimization problem; 2) Providing a reasonable interval for the optimal value; and 3) Performing proper benchmarks of the approximate suboptimal solution. The basic contributions of the thesis are summarized as follows:

\begin{enumerate}
  \item  We formulate a resource allocation problem to minimize the second-order difference between the achievable capacity and user request traffic in R-NOMA-BF system. We jointly optimize power, BF vector and LEO satellite-cell alignment factor to obtain  users' satisfaction. In the power-constrained case, we give constraints on the NOMA factor and BF vector. Further, we analyze the feasibility to match LEO satellite-cell and NP-hard.
  \item We design the Doppler shift LEO satellite-cell assisted monotonic programming of NOMA with BF vector (D-mNOMA-BF) algorithm. More specifically, the travel angles between relay satellite and LEO satellites are effectively measured as Doppler shifts, where users and cells are matched based on the measured results. BF vector optimization utilizes the singular value decomposition (SVD) algorithm to first isolate the interference at the cell level. Then, we design the monotone approximation optimization scheme to deal with the non-convexity about the objective function for NOMA power variables.
  \item We design an ant colony pathfinding based NOMA exponential cone programming with BF vector (A-eNOMA-BF) algorithm. Relay satellite is able to cover and deal with large-scale LEO satellites with cell matching data in real time by the ant colony algorithm. Compared with D-mNOMA-BF, this algorithm enables global planning of complex dynamic path information of LEO satellites. Furthermore, the NOMA-based exponential programming  scheme has higher accuracy and lower complexity.
  \item We further simulate the proposed NOMA-based algorithms superior to OMA for traffic fitting in R-NOMA-BF system. It is shown that A-eNOMA-BF algorithm satisfies the demand of users better than D-mNOMA-BF. We also verified two compromise algorithms, i.e., D-eNOMA-BF and A-mNOMA-BF. We compare the performance of different polarization and single-beam schemes. The  impact of users' satisfaction  on  ipSIC is taken into consideration.
\end{enumerate}
 \subsection{Organization and Notations}
 The rest of this paper is organized as follows. In Section II,  R-NOMA-BF  system is presented, where the ground users exist in cell aligned LEO satellite. In Section III, we formulate an optimization problem for power-constrained and LEO satellite-cell matching and discuss NP-hard. In Section IV, we provide two effective algorithms in  R-NOMA-BF system. Section V presents numerical results to verify the superiority of proposed algorithms, and concluded in Section VI.

The key symbols in this paper are elaborated as follows: The operator $\left| \cdot \right|$ represents the absolute value of a complex number. $\left\| {\cdot} \right\|$ indicates the square of the norm. ${\left(  \cdot  \right)^H}$ means the conjugate transpose of the matrix; ${\left(  \cdot  \right)^T}$ denotes transpose operations.
\section{System Model}\label{System Model}
As shown in Fig. 1, we consider a dynamic R-NOMA-BF communication system, where ground  users receive information from $M$ LEO satellites. The relay satellite is viewed as a geostationary earth orbit satellite to forward the gateway-LEO satellites or inter-LEO satellites information by measuring and tracking LEO satellites \cite{Litian2021,Wang2023feng}. Assume that a LEO satellite serves $N$ non-orthogonal ground users, the LEO satellites and users are equipped with multiple antennas and single antenna, respectively. More specifically, all users are served in the same satellite index and occupy the same bandwidth. Besides, the coverage of  satellite constellation is divided into $K$ cells. Assuming that  the set of time slots for the total period is defined as ${\cal T}$, i.e., ${\cal T} = \left[ {{t_1}...{t_k}...{t_K}} \right]$ and $\left| {\cal T} \right| = K$. Each LEO satellite serves a cell without duplication within per time slot. It is further explained that LEO satellites cover all the cells once in ${\cal T}$ time slots according to the optimized order. It is worth noting that the satellites mentioned later are referred to LEO satellites. \textcolor[rgb]{0.00,0.00,1.00}{The relay satellite can coordinate multiple transmission points to mitigate the effects of outdated CSI by providing more robust coverage. At the same time, handover and coordination of different satellites help to maintain accurate CSI in the LEO constellation.}

Considering the insufficient stability of inter-LEO satellite links, relay satellite assists LEO constellation to realize communication with users. The relay satellite could also solve the situation, where satellites are not visible along with ground stations. To collect the transmission signals and doppler shifts of high-speed mobile satellite satellites, relay satellite applies high-speed modems and multi-band tracking antennas to achieve the coverage and tracking of satellites. The relay satellite improves the emergency response capability of LEO constellation, which gains more warning time for responding to major natural disasters. As LEO satellites pass over unmanned ground stations, relay satellite transmits a large amount of user data in real time\footnote{Similar relay satellite assisted LEO satellites have many realistic application systems, such as the tracking and data relay satellite system, the  TianLian I 05 system and the data relay test satellite system.}. The relay satellite measures the status information of LEO constellation with multi-band tracking antenna and capture tracking receiver. The communication satellite payload is responsible for the communication between the LEO satellites, including the radio frequency system, antenna and so on. The satellite computer performs data processing and control tasks on board satellites \cite{Israel1993}. Meanwhile, the satellite transmitter adopts the phase and amplitude of the control signal to manage the direction and strength.  NOMA-BF has ability to improve system spectral efficiency  without increasing the antenna power for complex channel environments. NOMA-BF fully utilizes both spatial dimension and power domain resources. Specifically, BF technology can isolate inter-beam interference significantly, while NOMA compensates the spilling over inter-beam interference. Compared to BF, NOMA-BF saves satellite power consumption to distinguish users by different power. Each target terminal cannot fully realize spatial diversity simply by BF. NOMA-BF can multiplex power in the same direction, which reduces the cost of large-scale antennas.
\begin{figure}[h!]
\begin{minipage}[h!]{\linewidth}
\centering
\hspace{0cm}
\includegraphics[width=3.7in]{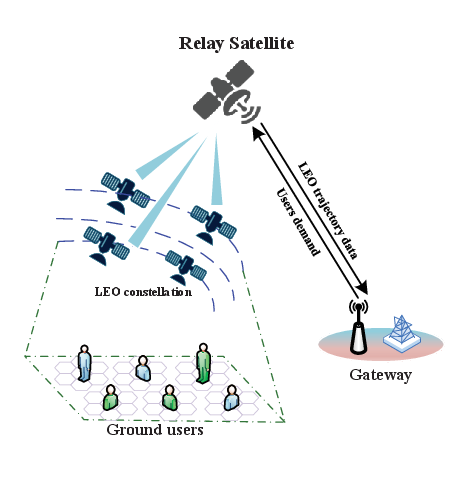}
 \caption{A relay satellite assisted LEO constellation  NOMA and BF communication system.}
\label{Fig. 1}
\end{minipage}
\end{figure}

In R-NOMA-BF system, satellites are matched with different cells, where each cell contains non-orthogonal users served for the satellites. Let ${\bf{H}} = \left[ {{{\bf{h}}_1} \cdots {{\bf{h}}_i} \cdots {{\bf{h}}_j} \cdots {{\bf{h}}_N}} \right]$ denote the downlink channel from satellite $m$ to ground users. The effective gains are ordered as ${\left\| {{{\bf{h}}_i}} \right\|^2} > {\left\| {{{\bf{h}}_j}} \right\|^2}$, where ${{\bf{h}}_i} = \left[ {{h_{i1}} \cdots {h_{il}} \cdots {h_{iL}}} \right]$. $L$ and $l$ represent the  number of transmission antennas and the corresponding transmission antenna index allocated to users, respectively. Considering the free-space path loss in  R-NOMA-BF system, the channel of the $l$-th antenna of user $i$ is expressed as \cite{Lin2023}
\begin{align}\label{eq1}
h_{il}^{{t_k}} = \vartheta   \sqrt {{G_r}} \nu ({d_{tk}})\sqrt {{G_t}\left( \varphi  \right)} {\bf{a}}\left( {\theta _i^{{t_k}}} \right),
\end{align}
where $\nu ({d_{tk}}) = \sqrt {{{[\lambda /(4\pi {d_{tk}})]}^2}} $ represents the large-scale fading coefficient. $\vartheta $ is defined as the rain attenuation factor, where $\vartheta$  grows by 0.01 dB per 1 km improvement of distance \cite{Kanellopoulos,3GPP171}. ${d_{tk}}$ is the distance between the satellite and the ground user at $t_k$ time slot, and $\lambda$ is defined as the wavelength. {Without loss of generality, the satellite base station is equipped with a uniform linear array. ${\rm{a}}\left( {\theta _i^{{t_k}}} \right) = \frac{1}{{\sqrt L }}{\left[ {\begin{array}{*{20}{l}}
1&{{e^{ - j\frac{{2\pi s}}{\lambda }\sin \theta _i^{{t_k}}}}}& \ldots &{{e^{ - j(L - 1)\frac{{2\pi s}}{\lambda }\sin \theta _i^{{t_k}}}}}
\end{array}} \right]^T}$ denotes the array response vector in the $i$-th user direction. ${\theta _i^{{t_k}}}$ is the angle-of-departure of the user $i$ at $t_k$ time slot. ${d_{tk}}$ and ${\theta _i^{{t_k}}}$ vary with LEO satellite movement and LEO satellite-cell matching. $s$ represents the neighboring antenna distance. ${{G_r}}$ represents the receive gain of all user antennas. ${{G_t}\left( \varphi  \right)}$ is the gain of the satellites transmitting antenna, which can be denoted as
\begin{align}\label{eq2}
G_t(\varphi)= \begin{cases}G_{\max }, & \varphi=0^{\circ} \\ G_{\max }\left|\frac{4 J_1(2 \pi a \sin \varphi / \lambda)}{2 \pi a \sin \varphi / \lambda}\right|^2, & 0^{\circ}<|\varphi| \leq 90^{\circ}\end{cases}
\end{align}
where $J_1(\cdot)$ is a Bessel function of the first class of the first order. $G_{\max }$ is the maximum transmitting antenna gain, which is indicated as ${G_{\max }} = 10 \times {\log _{10}}\left[ {{{\left( {\frac{{2\pi a}}{\lambda }} \right)}^2}} \right]$. $\varphi $ refers to the off-axis angle. $a$ is the radius of the antenna circular aperture. For single cell-aligned LEO satellite scenario, the signal of the satellite base station overlay is transmitted to the covered terrestrial $N$ users. On the basis of this, the received signal of a randomly selected user $i$ in the area covered by satellite $m$ is given by
\begin{align}\label{eq3}
y_i=\mathbf{\bf{h}}_i^{H} \mathbf{w}_i x_i+ \! \! \! \! \sum_{i^{\prime} \in \mathcal{N} \backslash\{i\}} \! \! \! \! \mathbf{\bf{h}}_i^{H} \mathbf{w}_{i^{\prime}} x_{i^{\prime}}+\sigma^2,
\end{align}
where ${\cal N}$ denotes the set of users covered via the LEO satellite. ${{\bf{w}}_i}$ represents the BF vector of user $i$. $\sigma^2 \sim \mathcal{C N}\left(0, N_0\right)$ is the Gaussian white noise. By utilizing SIC  scheme, the user of poor channel gain less than user $i$ are deleted. As a consequence, the signal-plus-interference-to-noise ratio (SINR) of user $i$ is expressed as \cite{Y2018xinwei}
\begin{align}\label{eq4}
{\gamma _i} = \frac{{{{\left| {{\bf{h}}_i^H{{\bf{w}}_i}} \right|}^2}{p_i}}}{{\sum\limits_{j \in {\cal N}\backslash \{ i\} ,j < i} \! \! \! \! \! \!  {{{\left| {{\bf{h}}_i^H{{\bf{w}}_j}} \right|}^2}{p_j}}  + {\sigma ^2}}},
\end{align}
where ${{p_i}}$ denotes the transmit power from corresponding beam to user $i$ . $ \! \! \! \! \sum\limits_{j \in {\cal N}\backslash \{ i\} ,j < i}\! \! \! \! \! \! \! \! {\left| {{\bf{h}}_i^{\it{H}}{{\bf{w}}_j}} \right|^2}{p_j}$ indicates intra-satellites interference. Assume that the inter-satellites interference is negligible. Thus, the achievable rate of user $i$ in cell $k$ covered by satellite $m$ at time slot ${{t_k}}$ can be expressed as
\begin{align}\label{eq5}
{R_i}{\rm{ = }}{\mu _{m{{t_k}}}}{B_{m{{t_k}}}}{\log _2}\left( {1 + {\gamma _i}} \right),
\end{align}
where ${\mu _{m{{t_k}}}}$ is the index whether  cell $k$ used by satellite $m$ or not at time slot ${{t_k}}$. ${B_{m{t_k}}}$ is expressed as the single carrier bandwidth of the satellite. Satellite trajectory data and Doppler shift values measured by relay satellite are brought into the mathematical model in subsequent algorithms.
\section{ PROBLEM FORMULATION AND ANALYSIS}

\subsection{Problem Formulation}
In the subsection, we formulate the joint optimization problem based on LEO satellite-cell assignment, power factor and BF vector for R-NOMA-BF system. To evaluate user satisfaction, $\sum {{{\left( {{R_i} - {D_i}} \right)}^2}} $ is applied to calculate the gap between the requested traffic and the available capacity. ${{D_i}}$ is defined as the requested traffic demand for user $i$. Furthermore, the optimization problem follows the conditions:
\begin{enumerate}
\item \emph{LEO satellite-cell assignment constraint:} Relay satellite determines the binary variables ${\mu _{m{t_k}}} = \left\{ {\left. {0,1} \right\}} \right.$ achieving alignment of LEO satellite-cell according to the Doppler shift and moving track of collaboration satellites. $\mu_{m t_k}=1$ is defined as LEO satellite $m$ successfully serves cell $m$ at moment $t_k$, and vice versa $\mu_{m t_k}=0$. Then ${\mu _{m{t_k}}}$  should satisfies：
    \begin{align}\label{LEO satellite-cell assignment constraint}
    \sum\limits_{m \in {\cal M}} {\mu _{m{t_k}}}  = M,\forall k \in {V_f},\forall {t_k} \in {\cal T},
    \end{align}
where ${V_f}$ represents the set of cells containing  all LEO satellites expected to be matched, e.g., the satellites  easily overcome with Doppler shift or stronger channels generated by distance. (6) denotes that the sum of cells corresponding to $V_f$ chosen by the satellites is $M$ at any moment $t_k$. There is no overlap among $V_f$ obtained by each satellite, which ensures that satellites cover separate cells at the same moment.
\item \emph{Power allocation factor constraint:} LEO satellite individually serves one cell after selection per time slots, so the power of users is bounded by the power on star. Each satellite-served users performs downlink NOMA protocols within a cell, so that it can be expressed as：
    \begin{align}\label{Power allocation factor constraint,eq1}
    \sum\limits_{i \in {\cal N}} {{p_i} \le P_s,\forall i \in } {\cal N},
    \end{align}
    \begin{align}\label{Power allocation factor constraint,eq2}
    {p_i} \le {\mu _{m{t_k}}}P_s,\forall i \in {\cal N},\forall k \in {\cal K},\forall {t_k} \in {\cal T},
    \end{align}
where $P_s$ is defined as the maximum power of the satellite base station transmitting. ${\cal K}$ represents the cell set of the determined satellite service. The sum power of users covered does not exceed $P_s$ by LEO satellite and cell alignment. (7) illustrates that the total power of all users cannot exceed $P_s$ in the cell. Meanwhile, (8) confines that the power of non-matching users is 0. The power of aligned users is also less than or equal to $P_s$.
\item  \emph{BF vector constraint:} BF realizes the effect of energy concentration by adjusting the amplitude and phase of multiple transmitting antennas. The superimposed energy in the user direction does not exceed the allocated user power. To satisfy the limitation of overall transmit power, the BF vector is constrained in different directions that can be expressed as:
    \begin{align}\label{Beamforming vector constraint}
    {\left|| {{{\bf{w}}_i}} \right||^2} \le {p_i},\forall i \in {\cal N},
    \end{align}
    where each beam consists of $L$ antennas. (9) ensures that the sum of the power in different directions should not exceed the user power $p_i$.
\end{enumerate}

According to above-mentioned constraints, the problem of minimizing the gap in traffic is formulated as
   \begin{align}\label{P0}
{\rm{   }}{{\cal P}_0}:&\mathop {\min }\limits_{{{\mu _{m{t_k}}}},{p_i},{{\bf{w}}_i}} \! \! \! \! \sum\limits_{k \in {\cal K},i \in {\cal N}} {{{\left( {{R_i} - {D_i}} \right)}^2}} \tag{10a}\nonumber\\
&{\rm{s}}{\rm{.t}}{\rm{. \quad   constraints}}\left( 6 \right) - \left( 9 \right),\tag{10b}\nonumber\\
{\rm{        }} &\ \qquad {R_i} \ge R_i^{\min },\forall i \in {\cal N},\tag{10c}\nonumber
    \nonumber\end{align}
 where $R_i^{\min }$ is defined as the minimum transmission rate of the user. (10c) ensures that user $i$ satisfies the minimum achievable rate to guarantee the fairness of the set of users. As a further development, the optimization variables of ${{\cal P}_0}$ interact with each other, thus it is difficult to jointly optimize these components. Due to the coupling between the optimization variables of equations (10a) and (10c), the objective function is non-convex with respect to $\mu _{m{t_k}}$, $p_i$, and ${{\bf{w}}_i}$. LEO satellites serving diverse cells affect the power allocation according to (7). Besides, the feedback in NOMA optimization scheme also leads to different design of LEO satellite-cell through (8). It is worth noting that the BF vector of (9) determines the optimization result by controlling the signal phase and amplitude without affecting the resource allocation. Consequently the variable ${{\bf{w}}_i}$ can be separately discussed to maximize the upper bound of user satisfaction.
 \subsection{Complexity analysis of ${{\cal P}_0}$}
 By the virtue of previous descriptions, ${{\cal P}_0}$ belongs to a MINCP problem originated from formulating nonlinear and non-convex functions. As a consequence, the feasibility of ${{\cal P}_0}$ is first checked after fixing ${{{\bf{w}}_i}}$\footnote{The BF vector is considered alone as the optimization variable that raises the upper bound on user satisfaction. ${{\bf{w}}_i}$ and other variables break down the coupling. As many joint optimization algorithms, ${{\bf{w}}_i}$ is first defined and fixed to an average value of the energy for different antenna directions. Then the ${{\cal P}_0}$ feasibility with respect to $\mu _{m{t_k}}$ and $p_i$ is analyzed by the fixed ${{\bf{w}}_i}$.}. To test whether a feasible solution exists for ${{\cal P}_0}$, it is expressed as a true-false problem. Since solving ${{\cal P}_0}$ directly is more challenging, the decision version can be obtained more conveniently. Assume that the decision version of ${{\cal P}_0}$ is NP-complete, the formulated optimization is NP-hard \cite{Deng2008}.
 \begin{proposition}\label{Theorem:NP-complete.}
The feasibility check problem for ${{\cal P}_0}$ is NP-complete.

\begin{proof}\label{Proof:NP-complete.}
A simplified NP-complete problem of ${{\cal P}_0}$ is formulated through the well-known three-dimensional matching problem.  To present straightforward results and analysis, it is further emphasized that the total number of time slots window and cells collection are regarded as the same, i.e., ${\cal N} = {\cal T}$. The set of cells is divided equally into two subsets depending on the number of units, ${{\cal K}_a}$ and  ${{\cal K}_b}$. Let ${{\cal K}_a}$ and  ${{\cal K}_b}$ have no identical elements while satisfying ${{\cal K}_a} \cup {{\cal K}_b} = {\cal K}$ and $\left| {{{\cal K}_a}} \right| = \left| {{{\cal K}_b}} \right| = \frac{K}{2}$. Therefore, the channel conditions with $k \in {{\cal K}_\varphi },\forall \varphi  \in \left\{ {a,b} \right\}$ are expressed as:
\begin{align}\label{P0}
{\left| {{h_{k'k}}} \right|^2} = \left\{ \begin{array}{l}
1 + \eta , \ {\rm{  }}k' = k,\\
1 + \frac{\eta }{2}, {\rm{ }}k' \in {{\cal K}_\varphi }{\rm{ }},\\
\eta ,\quad \quad \!{\rm{       }}k' \in {{\cal K}_{\varphi '}}{\rm{   }},
\end{array} \right.\tag{11}\nonumber
\nonumber\end{align}
where $0 \le \eta  \le {2^{\frac{1}{K}}} - 1$. It is assumed that the parameters are simple on an executable basis, where $P_s = 1$, $M = \frac{K}{2}$, ${\sigma ^2} = \eta $, $R_k^{\min } = 1$ and ${D_k} \ge R_k^{\min }$. At this moment, the hypothetical three-dimensional matching problem is considered to be present in the set $\Omega  \subset {\cal M} \times {{\cal K}_a} \times {{\cal K}_b}$. ${\cal M}$ is naturally defined as the group of satellites. Two points $\left( {{m},{k_a},{k_b}} \right)$  and $\left( {m',{k_{a'}},{k_{b'}}} \right)$ of $\Omega $ are arbitrarily chosen to satisfy $m \ne m'$, ${k_a} \ne {k_{a'}}$ and ${k_b} \ne {k_{b'}}$. As a result, the collection of cells with the same satellite index is necessarily in separate subsets. The achievable rate in satellite $m$-cell $k$ is calculated as ${\log _2}\left( {1 + \frac{{{{\left| {{h_{kk}}} \right|}^2}P_s}}{{{{\left| {{h_{k'k}}} \right|}^2}P_s + {\sigma ^2}}}} \right) = {\log _2}\left( {1 + \frac{{1 + \eta }}{{\eta  + \eta }}} \right) > 1 = R_k^{\min }$. The result of the above calculation just makes (7) hold. Overall, solving the formulated problem is feasible and the proof is completed.
\end{proof}
\end{proposition}
\begin{corollary}\label{NP-hard}
The three-dimensional matching problem was determined to be true. If two cells are observed on the same satellite index in the common subset, the rates of cells are given: ${\log _2}\left( {1 + \frac{{{{\left| {{h_{kk}}} \right|}^2}P_s}}{{{{\left| {{h_{k'k}}} \right|}^2}P_s + {\sigma ^2}}}} \right) = {\log _2}\left( {1 + \frac{{1 + \eta }}{{1 + \frac{\eta }{2} + \eta }}} \right) < {\log _2}(1 + 1) = R_k^{\min }$. The conclusion is obviously contrary to the constraint of (10c). Consequently, the cells at the same satellite have to come from different subsets. ${{\cal P}_0}$ is NP-hard on the basis of the feasibility check being 'yes'.
\end{corollary}
\begin{remark}
As various cell $k$ matches different LEO satellites $m$ based on (6), the cell must exceed or equal to $R_k^{\min }$ under the condition of (7) and (8). Hence simplified example satisfying (6)-(8) cannot violate the constraints of (10c), which suggests that $\mathcal{P}_0$ has a feasible solution.
\end{remark}
It is important to note that ${{\cal P}_0}$ remains non-convex after fixing ${{{\bf{w}}_i}}$ and refining the binary variable. In spite of the multidimensional variables interfering with each other, the formulated problem still requires approximation and transformation of an acceptable solution.
\section{An effective joint optimization algorithm proposed}
In this section, the D-mNOMA-BF algorithm stepwise expansion is discussed in  R-NOMA-BF system, which performs a coupled solution cycle after iterating the parts. In light of the above work, the complexity and pseudo-code of this algorithm are listed. In  D-mNOMA-BF algorithm, ${\mu _{m{t_k}}}$ and ${{p_i}}$ are jointly optimized and iterated step by step after fixing the BF vector, while the problem about the ${{{\bf{w}}_i}}$ construction approximation is further solved.
\subsection{Doppler shift-based LEO satellite-cell matching strategy}
With the aid of the function of tracking and measuring LEO satellite for the relay satellite, an LEO satellite-cell aligning strategy involving doppler shift threshold  is designed. As LEO satellite travels along a certain direction at the constant rate ${{v_0}}$, it causes a change of wavelength $\lambda $ owing to the distance gap in the signal transmission. The wavelength of variation is defined as $\bar \lambda {\rm{ = }}\frac{{{c^2}}}{{(c \pm {v_0}\cos \alpha ){f_c}}}$, where ${{f_c}}$ and $c$ are the original carrier frequency and the speed of light constant, respectively. $\alpha $ represents the angle between the direction of LEO satellite movement and relay satellite link. In addition, the Doppler frequency shift denoted by $f' = \frac{{{f_c}{v_0}\cos \alpha }}{c}$. Let ${\Gamma _f}$ denotes the threshold value of new frequency. Hence $M$ LEO satellites match all the cells satisfying the condition of ${{f'_m}} \le {\Gamma _f}$ in the time slot ${t_k}$. If the checked cells have no overlap, the set of LEO satellite-cell is expressed as ${V_f}\left( {{M_{{t_k}}},K} \right)$. Otherwise, ${m_1}$ and ${m_2}$ match both ${k_\varepsilon }$ by the means of satisfying the threshold condition, while either ${m_1}$ or ${m_2}$ of the larger frequency shift drops ${k_\varepsilon }$. The remaining LEO satellites will pick the cells from ${{V'_f}}\left( {{M_{{t_k}}},\tilde K} \right)$  to fill the empty spots. Next, the residual collection of cells $\{ \tilde K\} $ is the time slot ${t_k} + 1$ fitting object, i.e. ${M_{{t_k} + 1}} \in \{ \tilde K\} $. By exploiting the above sufficient logical reasoning, the ${V_f}$ of constraint (6) completes the connecting according to the Doppler frequency shift threshold.
\subsection{Monotonic programming of power allocation}
The monotonic programming of  power allocation is applied after fixing ${{{\bf{w}}_i}}$. The assumption is that the power of all users belongs to non-negative orthogonal values. On the basis of the above assumption, (5) as a difference function is rewritten as
\begin{align}\label{12}
{R_i} \nonumber
& = {\mu _{m{t_k}}}{B_{m{t_k}}}{\log _2}\left( {1 + \gamma \left( {{p_i}} \right)} \right) \\ \nonumber
&= {\mu _{m{t_k}}}  \left( {{\xi ^ + }\left( {{p_i}} \right) - {\xi ^ - }\left( {{p_i}} \right)} \right).\tag{12}\nonumber
\nonumber\end{align}
The assumption generally holds for typical expression (4). Then the $\xi $ of difference function can be given by
\begin{align}\label{13}
{\xi ^ + }\left( {{p_i}} \right) = {B_{m{t_k}}}{\log _2}\left( {{\sigma ^2} + {{\left| {{\bf{h}}_i^{{H}}{{\bf{w}}_i}} \right|}^2}{p_i} \! \!+\! \! \! \! \sum\limits_{\mathop {j \in {\cal N}\backslash \{ i\} }\limits_{j < i} } {{{\left| {{\bf{h}}_j^{{H}}{{\bf{w}}_j}} \right|}^2}{p_j}} } \right),\tag{13}\nonumber
\nonumber\end{align}
and
\begin{align}\label{14}
{\xi ^ - }\left( {{p_i}} \right) = {B_{m{t_k}}}{\log _2}\left( {{\sigma ^2} +\! \! \! \! \sum\limits_{\mathop {j \in {\cal N}\backslash \{ i\} }\limits_{j < i} } \! \! \! \! {{{\left| {{\bf{h}}_j^{{H}}{{\bf{w}}_j}} \right|}^2}{p_j}} } \right),\tag{14}\nonumber
\nonumber\end{align}
respectively. Furthermore, a monotonic function $f$ is defined that the real vector space maps to the associated real matrix space, where $q \ge z$, then $f\left( q \right) \ge f\left( z \right)$. ${\xi ^ + }\left( {{p_i}} \right)$ and ${\xi ^ - }\left( {{p_i}} \right)$ are considered as a programming of satisfied the monotonic function $f$. It is worth mentioning that ${\xi ^ + }\left( {{p_i}} \right)$ and ${\xi ^ - }\left( {{p_i}} \right)$ are not distinguished from concave and convexity. At the same time, both the objective function and constraints of achievable rate can be rewritten in the form of monotonic programming function. Consequently, ${{\cal P}_0}$ is equivalently formulated as
\begin{align}\label{15}
{{\cal P}_1}:  \tag{15a}
&\mathop {\min }\limits_{{p_i}} {\sum\limits_{k \in {\cal K},i \in {\cal N}} {\left( {\left( {{\xi ^ + }\left( {{p_i}} \right) - {\xi ^ - }\left( {{p_i}} \right)} \right) - {D_i}} \right)} ^2}, \\
&{\rm{s}}{\rm{.t}}  {\rm{.          \qquad    constraints}}\left( 7 \right), \left( 8 \right), \left( {{\rm{10c}}} \right), \tag{15b}\nonumber
\end{align}
where ${\xi ^ + }\left( {{p_i}} \right)$ and ${\xi ^ - }\left( {{p_i}} \right)$ grow positively with ${{p_i}}$. Since the optimization function is a quadratic difference function, the monotonicity of ${{\cal P}_1}$ depends on the variation of ${{D_i}}$ value. To avoid the waste of payload on the satellites and select the optimal terminal, ${R_i} \le {D_i}$ has been proved in \cite{Wang2022A}. Therefore, ${{\cal P}_1}$ can be approximated as
\begin{align}\label{16}
{{{\cal P}'_1}}:\mathop {\min }\limits_{{p_i}} {\left( {{\mu _{m{t_k}}}\left( {\sum\limits_{\mathop {k \in {\cal K}}\limits_{i \in {\cal N}} } {{\xi ^ + }\left( {{p_i}} \right)}  - \sum\limits_{\mathop {k \in {\cal K}}\limits_{i \in {\cal N}} } {{\xi ^ - }\left( {{p_i}} \right)} } \right) - {D_i}} \right)^2}. \tag{16}\nonumber
\end{align}
As a further development, the auxiliary variable $\partial  = \! \! \! \! \sum\limits_{k \in {\cal K},i \in {\cal N}} \! \! \! {{\xi ^ - }\left( {{p_{\max ,i}}} \right) - {\xi ^ - }\left( {{p_i}} \right)}  $ is introduced, where ${p_{\max ,i}}$ is the maximum power available to user $i$. The optimization problem is reformulated as
\begin{align}\label{17}
{{\cal P}_2}: \nonumber
&\mathop {\min }\limits_{\partial ,{p_i}} {\left( {{\mu _{m{t_k}}}\left( {\sum\limits_{\mathop {k \in {\cal K}}\limits_{i \in {\cal N}} } {{\xi ^ + }\left( {{p_i}} \right)}  + \partial } \right) - {D_i}} \right)^2}, \tag{17a} \nonumber \\
&{\rm{s}}{\rm{.t}} {\rm{.    \qquad \quad   constraints}}\left( 7 \right),\left( 8 \right),\left( {{\rm{10c}}} \right),\nonumber \tag{17b} \\ \nonumber
& \qquad \qquad  \left\{ {\partial ,{p_i}} \right\} \in {\cal G}, \nonumber \tag{17c}
\end{align}
where
\begin{align}\label{18}
{\cal G} = \left\{ {\begin{array}{*{20}{c}}
{0 \le \partial  + {\xi ^ - }\left( {{p_i}} \right) \le {\xi ^ - }\left( {{p_{\max ,i}}} \right),}\\
{0 \le \partial  \le {\xi ^ - }\left( {{p_{\max ,i}}} \right) - {\xi ^ - }\left( {{p_{0,i}}} \right).}
\end{array}} \right. \nonumber \tag{18}
\end{align}
Despite the fact that ${{\cal P}_2}$ has refined the monotonicity of optimization function, the constraint (10c) is not continuously monotonic. In order to solve the above case, (13) and (14) are substituted into (10c) and valid constraint is obtained. As a result, the final optimization problem iteration for power is
\begin{align}\label{19}
{{\cal P}_3}:
&\mathop {\min }\limits_{\partial ,{p_i}} {\left( {{\mu _{m{t_k}}}\left( {\sum\limits_{\mathop {k \in {\cal K}}\limits_{i \in {\cal N}} } {{\xi ^ + }\left( {{p_i}} \right)}  + \partial } \right) - {D_i}} \right)^2},\tag{19a} \nonumber \\
& {\rm{s}}{\rm{.t}}{\rm{. \qquad \quad constraints}}\left( 7 \right),\left( 8 \right), \tag{19b} \nonumber \\
&\qquad \qquad \left\{ {\partial ,{p_i}} \right\} \in {\cal G}, \tag{19c} \nonumber \\
&\qquad \qquad \ {\xi ^ + }\left( {{p_i}} \right) + \partial  \ge R_i^{\min }. \tag{19d} \nonumber
\end{align}
The following change comes with
\begin{align}\label{20}
{\cal G} = \left\{ {\begin{array}{*{20}{c}}
{0 \le \partial  + {\xi ^ - }\left( {{p_i}} \right) \le {\xi ^ - }\left( {{p_{\max ,i}}} \right) - R_i^{\min },}\\
\begin{array}{l}
\! \! \! \! \! \! \! \! \! \! 0 \le \partial  \le {\xi ^ - }\left( {{p_{\max ,i}}} \right) - {\xi ^ - }\left( {{p_{0,i}}} \right),\\
\! \! \! \! \! \! \! \! \! \! {\xi ^ + }\left( {{p_i}} \right) + \partial  \ge {\xi ^ - }\left( {{p_{\max ,i}}} \right) - R_i^{\min }.
\end{array}
\end{array}} \right. \tag{20} \nonumber
\end{align}
\subsection{BF vector design}
For complex satellite-ground networks, the power optimization not only maximizes the utilization of LEO constellation payload, but also enhances user satisfaction. Since the BF applies a combination of antennas and digital signal processing techniques, it falls outside of a resource loads allocation. By exploiting the optimized ${{\mu _{m{t_k}}}}$ and ${{p_i}}$, we approximate the problem of minimizing traffic difference as maximizing effective BF capacity. The objective problem is further given by
\begin{align}\label{Pw}
{{\cal P}_1}({\bf{w}}):
&\mathop {\max }\limits_{{{\bf{w}}_i}} {\left| {{\bf{h}}_i^{{H}}{{\bf{w}}_i}} \right|^2}, \tag{21a} \nonumber \\
{\rm{s}}{\rm{.t}}{\rm{. }} \ & {\left|| {{{\bf{w}}_i}} \right||^2} \le {p_i},\forall i \in {\cal N}. \tag{21b} \nonumber
\end{align}
To avoid complex inverse solving and uncertain estimation during BF, (21a) is subjected to SVD \cite{Liu2009}. From the point of view of maximizing the received SINR, the BF  vector  should be derived according to the optimality criterion as follows:
\begin{align}\label{Pw2}
{{\cal P}_2}({\bf{w}}):\mathop {\max }\limits_{{{\bf{w}}_i}} {{\bf{H}}_i}{{\bf{W}}_i}, \tag{22} \nonumber
\end{align}
where  ${\bf{H}} = {{\bf{h}}^{{H}}}{\bf{h}}$, ${\bf{W}} = {{\bf{w}}^{{H}}}{\bf{w}}$. ${\bf{w}}_i^*$ indicates the corresponding vector consisting of the $L$ largest eigenvalues of ${{\bf{W}}}$. Then, the SVD decomposition of the channel matrix ${\bf{H}}$ is performed and the associated left singular vector can be found.
\begin{algorithm}\label{alg:D-mNOMA-BF}
\caption{D-mNOMA-BF}
\hspace*{0.001in} {\bf Input:}\;$\Gamma_{f}$, $D_i$
\begin{algorithmic}
\STATE  Initialize: ${{\mu _{m{t_k}}}}$, $p_i$, ${{{\bf{w}}_i}}$\\
1:\hspace*{0.005in}  {\bf repeat} \\
2:\hspace*{0.02in}  Calculate the set of cells that satisfy ${{f'_m}} \le {\Gamma _f}$\\
\hspace*{0.15in} according to (6).\\
3:\hspace*{0.02in}   {\bf for} $a=1:M$\\
4:\hspace*{0.2in}  Find the minimum value of the sum of users satisfac-\\
\hspace*{0.33in} tion in $V_f$.\\
5:\hspace*{0.02in}   {\bf end for} \\
6:\hspace*{0.02in} Obtain an alignment strategy ${{\mu _{m{t_k}}}}$ for LEO satellite-cell\\
 \hspace*{0.15in} in the moment.\\
7:\hspace*{0.02in} {\bf for} $b=1:M$\\
8:\hspace*{0.2in}  The left singular vector of the channel matrix is\\
\hspace*{0.33in} acquired on the basis of (22) with SVD.\\
9:\hspace*{0.2in}  Gain the optimized ${\bf{w}}_i^*$.\\
11:\hspace*{0.2in}  {\bf repeat} \\
12:\hspace*{0.35in}  Solve ${{\cal P}_3}$ under the condition of (20).\\
13:\hspace*{0.35in}  Get the optimized $p_i$.\\
14:\hspace*{0.2in} {\bf until convergence} \\
15:\hspace*{0.2in}  Calculate $R_i$ by (4) and (5).\\
16:\hspace*{0.02in}   {\bf end for} \\
17:\hspace*{0.005in} {\bf until $\forall i \in \mathcal{N}, \boldsymbol{R}_{i_{-} \text {temp } } \! \! \! =\boldsymbol{R}_i$}. \\
\STATE Obtain the optimal ${{\mu _{m{t_k}}}}$, $p_i$, ${{\bf{w}}_i}$.\\
\end{algorithmic}
\hspace*{0.001in} {\bf Output:}\;$R_i$\\
\end{algorithm}

As mentioned above, the proposed D-mNOMA-BF algorithm is summarized in Alg. 1. The main idea of D-mNOMA-BF algorithm consists of LEO satellite-cell alignment scheme with the aid of the Doppler shift threshold measured, and then jointly optimizing the power factor and BF vector. In particularly, lines 2 to 6 represent the Doppler shift constrained thresholds $\Gamma_{f}$  to design the LEO satellite-cell scheme. Lines 7 to 9 denote the utilization of SVD to yield left singular vector, i.e., ${{\bf{w}}_i}$. Finally, lines 11 to 15 indicate the power allocation factor for the application of monotonic planning in calculating NOMA. The D-mNOMA-BF algorithm first iterates ${{\bf{w}}_i}$ and $p_i$ so as to maximize the fit of user request traffic in the formulated objective problem, where the optimization of ${{\bf{w}}_i}$ allows the optimal value to be improved. $p_i$ makes the matching better. Then ${{\mu _{m{t_k}}}}$ depends on the Doppler shift generated from the high speed travel of the LEO satellite in R-NOMA-BF system. The cycle of two parts iteration until ${{R_i}}$ has a certain accuracy. Besides, the maximum number of times that all LEO satellites and cells are aligned and exchanged is defined as $MK \times \left( {1{\rm{ + }}K} \right)/2$. In D-mNOMA-BF algorithm, the interior point method is employed for solving ${{\cal P}_3}$ with the complexity of ${\cal O}\left( {\varpi \log \left( {\frac{1}{\vartheta }} \right)} \right)$, where $\varpi  > 0$, $\vartheta  > 0$. $\varpi $ and $\vartheta $ denote the parameter for self-concordant barrier and precision, respectively. The main complexity of SVD algorithm comes from matrix decomposition, which gives ${\cal O}\left( {{M^2}L + {L^2}M} \right)$. Hence, the complexity of Alg. 1 is ${\cal O}\left( {\left( {MK \times \left( {1{\rm{ + }}K} \right)/2} \right)\left( {{M^2}L + {L^2}M} \right) \times M\varpi \log \left( {\frac{1}{\vartheta }} \right)} \right)$ in  R-NOMA-BF system.

To gain more deep insights, the gap between the result of D-mNOMA-BF algorithm and the optimal value of ${{\cal P}_0}$ is three kinds:
\begin{itemize}
\item $\mathcal{P}_1(\mathbf{w})$ guarantees the increase of the optimal achievable rate, but it sacrifices the effectiveness of the traffic gap fitting.
\item The variable $\partial $ introduced by monotonic programming leads to a drop in traffic matching accuracy.
\item With respect to the user sets difference equation ${{{\cal P}'_1}}$ is actively divided into two parts, which appears as a cliff for the traffic fit.
\end{itemize}

\textcolor[rgb]{0.00,0.00,1.00}{To avoid the performance degradation caused by power optimization, ${{\tilde P}_s}$ is introduced instead of ${P_s}$ in equations (7) and (8). ${{\tilde P}_s}$ is a pre-value derived  form the request traffic of users, and slightly larger than ${P_s}$.}
\section{Exponential optimization of NOMA assisted ant colony satellite cooperative algorithm}
In this section, A-eNOMA-BF algorithm is presented to further increase users' satisfaction. It is demonstrated that ant colony pheromone has significant advantages for path optimization in the space-temporal decomposition. The exponential cone programming overcomes the non-convexity on the optimization problem. The A-eNOMA-BF algorithm retains the previous BF vector processing method.

\subsection{Ant colony algorithm with LEO satellite path planning}
The individual behavior of ants foraging has no statistical results, but large numbers of ants as a whole foraging is regarded as intelligent. More especially, colony sifts across different environments reaching the shortest path to food, where ants achieve information exchange by releasing pheromone. Simultaneously, ants have the ability to perceive pheromone and walk along paths with higher concentrations. After a certain time of feedback, all ants will take the fastest route on the food point. The LEO constellation and multiple cells together form a complex access network. Since LEO satellites exist in the different starting points, ant colony completes path closure as the base point in the cell. Consequently, LEO constellation can select the path of matching cells after intelligent optimization with the ant population.

The  A-eNOMA-BF algorithm is to deal with processed (6), that is, using the modified ant colony algorithm. ${{\cal M}_{{\rm{ant}}}}$ is defined as the primary ant colony. The coordinate origin of the relay satellite is ${u_0}$. Meanwhile, ${{\tilde u}_{mk}}$ is preset the original first point of the LEO satellite metaphor for ants, where each LEO satellite undoubtedly have different ones. Instead of multiple ground gateways, one relay satellite can measure the path of the LEO satellites as a whole. $c_m^{k \to k'} = {\left\{ {{u_0} \to {{\tilde u}_{mk}} \to {u_{mk'}}} \right\}_{{t_k}}}$  represents the possible lines of ant $m$ from ${u_0}$ to ${{u_{mk'}}}$. It is further stated that $\iota _m^{k \to k'}$ and $\ell _{kk'}^m$ are expressed as factors with route-based weights and pheromone concentrations, respectively. The probability of ants picking $c_m^{k \to k'}$ with ${{{\tilde u}_{mk}}}$ as its starting location is given as
\begin{align}\label{23}
P_m^{k \to k'}\left( \chi  \right) = \left\{ {\begin{array}{*{20}{c}}
{\frac{{{{\left( {\iota _m^{k \to k'}} \right)}^{{s_1}}} \times {{\left( {\ell _{kk'}^m} \right)}^{{s_2}}}}}{{\sum\limits_{\hat k \in {V_k}} {\left( {{{\left( {\iota _m^{\hat k \to k'}} \right)}^{{s_1}}} \times {{\left( {\ell _{\hat kk'}^m} \right)}^{{s_2}}}} \right)} }},\hat k \in {V_k}}\\
{ \ 0, \qquad \qquad \qquad \qquad \quad  \hat k \notin {V_k}}
\end{array}} \right. ,\tag{23} \nonumber
\end{align}
where $\chi $ is the number of iterations finding paths for ants. Also, ${{s_1}}$ and ${{s_2}}$ respectively indicate the critical degree order with respect to ${\iota _m^{k \to k'}}$ and $\ell _{kk'}^m$. Note that ${\hat k \in {V_k}}$ means the collection of accessible  cells  in the vicinity from satellite $m$. Once the colony accomplishes route mapping, the pheromone for $c_m^{k \to k'}$ is adjusted to
\begin{align}\label{24}
\ell _{kk'}^m\left( {\chi  + 1} \right) = \left( {1 - \tau } \right)\ell _{kk'}^m\left( \chi  \right) + \Delta \ell _{kk'}^m,\tag{24} \nonumber
\end{align}
where $\tau $ denotes the pheromone volatility ratio. $\ell _{kk'}^m$ represents the pheromone addition on $c_m^{k \to k'}$, e.g., $\Delta \ell _{kk'}^m = \!\!\!\!\!\!\!\sum\limits_{{m_a} \in {{\cal M}_{{\rm{ant}}}}} \!\!\!\!\! {\Delta \ell _{kk'}^{{m_a}}} $,  which is the pheromone released by ant ${{m_a}}$ within set ${{\cal M}_{{\rm{ant}}}}$. ${\Delta \ell _{kk'}^{{m_a}}}$ is positively correlated with ${R_{{t_k}}}$, where ${R_{{t_k}}}$ is the LEO satellite movement speed at time slot ${t_k}$.

Next, ${{\cal P}_0}$ optimization problem is iterated continuously according to the modified ant colony algorithm. The alignment strategy of ${V_f}$ is regarded as the link selection approach. Thus, the optimization problem is reformulated as
\begin{align}\label{25}
&{{\cal P}_0}\left( a \right):\mathop {\min }\limits_{{\mu _{m{t_k}}},{c_m}} \!\! \sum\limits_{k \in {\cal K},i \in {\cal N}} \!\!\!\!\! {{{\left( {{R_i} - {D_i}} \right)}^2}} \nonumber \tag{25a} \\
& {\rm{s}}{\rm{.t}}{\rm{. }}\sum\limits_{m \in {\cal M}} {{\mu _{m{t_k}}} = M,\forall k \in {V_f},\forall {t_k} \in {\cal T}}, \nonumber \tag{25b} \\
&\quad \ \  c_m^{k \to k'} \subset {V_f},\forall m \in {\cal M}. \nonumber \tag{25c}
\end{align}
${{\cal P}_0}\left( a \right)$ needs to jointly optimize ${{\mu _{m{t_k}}}}$ and ${{c_m}}$ for sufficient iterations. Furthermore, the threshold $\ell _{kk'}^{\max }$ limiting $\ell _{kk'}^m$ has to be introduced in the process of solving  $c_m^{k \to k'}$.
\subsection{Exponential Cone Programming for NOMA}
After obtaining the distribution of the coordinated LEO satellite-cell, NOMA power factor and  BF vector are jointly optimized once again. We then apply BF vector optimization idea in the previous section. Further providing valuable insights and fixing the optimized ${{{\bf{w}}_i}}$, regarding the optimization ${{p_i}}$ of ${{\cal P}_0}$ is a complex MINCP. If the optimization variable is replaced from ${{p_i}}$ to ${{R_i}}$, it would be an easy process for solving the objective function. However, due to the presence of the (10c) constraint, the above mentioned transformation has some difficulty. It is appropriate to bring the exponential cone programming into solving the constraint process. Assume that inter-beam interference between users concerning BF vector can be  negligible on satellites. The NOMA power factors for all users ordered on the basis of channel gain are expressed separately as
    \begin{align}\label{26}\nonumber
    &{p_1} = \left( {{2^{\frac{{{R_1}}}{{{B_{m{{t_k}}}}}}}} - 1} \right)\frac{{{\sigma ^2}}}{{{{\left| {{\bf{h}}_1^H{{\bf{w}}_1}} \right|}^2}}},\\\nonumber
    &{p_2} = \left( {{2^{\frac{{{R_2}}}{{{B_{m{{t_k}}}}}}}} - 1} \right)\left( {{p_1} + \frac{{{\sigma ^2}}}{{{{\left| {{\bf{h}}_2^H{{\bf{w}}_2}} \right|}^2}}}} \right),\\\nonumber
    &......\\\nonumber
    &{p_N} = \left( {{2^{\frac{{{R_N}}}{{{B_{m{{t_k}}}}}}}} - 1} \right)\left( {\sum\limits_{\hat i = 1}^{N - 1} {{p_{\hat i}}}  + \frac{{{\sigma ^2}}}{{{{\left| \nonumber {{\bf{h}}_N^H{{\bf{w}}_N}} \right|}^2}}}} \right).\tag{26}
    \end{align}
According to the user power consequences derived with approximate assumptions, (7) and (8) are recalculated as
    \begin{align}\label{11}
      \sum\limits_{i = 1}^N {\left( {\frac{{{\sigma ^2}}}{{{{\left| {{\bf{h}}_i^H{{\bf{w}}_i}} \right|}^2}}} - \frac{{{\sigma ^2}}}{{{{\left| {{\bf{h}}_{i - 1}^H{{\bf{w}}_{i - 1}}} \right|}^2}}}} \right)} {2^{\frac{{\sum\limits_{\hat i = i}^N {{R_{\hat i}}} }}{{{B_{m{{t_k}}}}}}}} - \frac{{{\sigma ^2}}}{{{{\left| {{\bf{h}}_N^H{{\bf{w}}_N}} \right|}^2}}} \le {P_s},\tag{27}\nonumber
    \end{align}
where let $\frac{{{\sigma ^2}}}{{{{\left| {{\bf{h}}_0^H{{\bf{w}}_0}} \right|}^2}}} = 0$. In A-eNOMA-BF algorithm, the optimization problem about power is reformulated as
\begin{align}\label{11}
&{{\cal P}_1}\left( a \right):\mathop {\min }\limits_{{R_i},{\mu _{m{t_k}}},{c_m}} \!\! \sum\limits_{k \in {\cal K},i \in {\cal N}} \!\!\! \! \! {{{\left( {{R_i} - {D_i}} \right)}^2}} \nonumber \tag{28a} \\
&{\rm{s}}{\rm{.t}}{\rm{.   \quad   constraints \ (25b),(25c),(27)}}.\nonumber \tag{28b}
\end{align}

\begin{algorithm}\label{alg:A-eNOMA-BF}
\caption{A-eNOMA-BF}
\hspace*{0.001in} {\bf Input:} $\ell _{kk'}^{\max }$, $D_i$, $\ell _{kk'}^m$, $\iota _m^{k \to k'}$, ${{{\tilde u}_{mk}}}$, ${{\cal M}_{{\rm{ant}}}}$, ${{s_1}}$, ${{s_2}}$, $\tau $\\ \vspace{-0.4cm}
\begin{algorithmic}
\STATE  Initialize: ${{\mu _{m{t_k}}}}$, $p_i$\\
1:\hspace*{0.005in} {\bf repeat}\\
2:\hspace*{0.02in} $\chi  = 1$, ${t_k} = 0$\\
3:\hspace*{0.15in} {\bf for} ${t_k} \le {t_K}$\\
4:\hspace*{0.2in}  Search for satellite accessible cell aggregation ${{V_k}}$.\\
5:\hspace*{0.2in}  {\bf while} ${{m_a} \in {{\cal M}_{{\rm{ant}}}}}$ {\bf do}\\
6:\hspace*{0.25in} Set the initial route weighting factor $\iota _m^{k \to k'}$.\\
7:\hspace*{0.25in} Initialize this cycle pheromone  $\ell _{kk'}^m$.\\
8:\hspace*{0.30in} {\bf for} $\ell _{kk'}^m \le \ell _{kk'}^{\max }$\\
9:\hspace*{0.35in} Define iteration $c_m^{k \to k'}$.\\
10:\hspace*{0.40in} {\bf for} $z = 1:M$\\
11:\hspace*{0.45in} Set the source point of each ant.\\
12:\hspace*{0.45in} Calculate $P_m^{k \to k'}$ in ${{V_k}}$ according to (23).\\
13:\hspace*{0.45in} Select the pheromone of the link by (24).\\
14:\hspace*{0.40in} {\bf end for}\\
15:\hspace*{0.30in} {\bf end for} \\
16:\hspace*{0.30in} Calculate the link $c_m^{k \to K}$ achievable capacity and  \\
\hspace*{0.48in} select the maximum value.\\
17:\hspace*{0.30in} Update the pheromone concentration at $c_m^{k \to K}$.\\
18:\hspace*{0.30in} $\chi  = \chi  + 1$.\\
19:\hspace*{0.30in}  Update the route weights $\iota _m^{k \to k'}$.\\
20:\hspace*{0.30in}  {\bf if} $t_k^{now} > {t_K}$ {\bf then}\\
21:\hspace*{0.40in}  {\bf break}\\
22:\hspace*{0.30in}  {\bf end if}\\
23:\hspace*{0.2in}  {\bf end while}\\
24:\hspace*{0.15in}  {\bf end for}\\
25:\hspace*{0.15in}  Calculate ${V_f}$ and give the LEO satellite-cell align-\\
\hspace*{0.35in} ment strategy ${{\mu _{m{t_k}}}}$.\\
26:\hspace*{0.15in}  Obtain the optimized ${{{\bf{w}}_i}}$ via Alg. 1.\\
27:\hspace*{0.2in}  {\bf repeat} \\
28:\hspace*{0.35in}  Calculate the NOMA user power for (26).\\
29:\hspace*{0.35in}  Solve (27) power constraints.\\
30:\hspace*{0.35in}  Optimize ${{p_i}}$ in ${{\cal P}_0}\left( a \right)$.\\
31:\hspace*{0.2in} {\bf until convergence} \\
32:\hspace*{0.2in}  Calculate $R_i$ by (4) and (5).\\
33:\hspace*{0.005in} {\bf until $\forall i \in \mathcal{N}, \boldsymbol{R}_{i_{-} \text {temp } }=\boldsymbol{R}_i$}. \\
\STATE Obtain the optimal ${{\mu _{m{t_k}}}}$, $p_i$, ${{{\bf{w}}_i}}$.\\
\end{algorithmic}
\hspace*{0.001in} {\bf Output:}\;$R_i$\\
\end{algorithm}

The flow of A-eNOMA-BF algorithm is summarized in Alg. 2, which includes three kinds, i.e., modified ant colony, NOMA with exponential cone programming and existing SVD algorithms on BF vector. In particular, lines 2 to 9 define key information such as the  pheromone concentrations  and route weights of ant colony algorithm. Lines 11 to 17 address the LEO satellite link access cells probability. Lines 18 to 25 indicate the acquired optimized ${{\mu _{m{t_k}}}}$ through pheromone iterations of ant colony. Moreover, BF vector resulting from Alg. 1 and NOMA power factor of  exponential cone are jointly optimized in the remaining fraction. We finally computed the overall iterated ${{R_i}}$ for ${{\cal P}_0}\left( a \right)$ in R-NOMA-BF system. Compared to D-mNOMA-BF algorithm, it has a globally integrated LEO satellite-cell strategy of making more accurate matching. Simultaneously, the exponential cone programming is easily on finding the minimum gap to realize the simplification constraint. In conclusion, the complexity of modified ant colony algorithm is calculated as ${\cal O}\left( {{t_K}\chi KM \cdot {{\cal M}_{{\rm{ant}}}}} \right)$. The overall complexity is given as ${\cal O}\left( {{J_{\max }}\left( {{M^2}L + {L^2}M} \right)M\varpi \log \left( {\frac{1}{\vartheta }} \right) \times \left( {{t_K}\chi KM \cdot {{\cal M}_{{\rm{ant}}}}} \right)} \right)$ in A-eNOMA-BF algorithm, where ${{J_{\max }}}$ is its maximum number of iterations. For A-eNOMA-BF algorithm, besides the error generated by SVD approximation value, ${{\cal P}_0}\left( a \right)$ ignores the interference of multiple antennas among users which also affects the optimal solution.
\section{NUMERICAL RESULTS}
\subsection{Simulation Parameter and Benchmarks}
This section provides the numerical results to evaluate the performance of the proposed algorithms in R-NOMA-BF system. The course of simulation still holds the same objective as $\sum\limits_{k \in {\cal K},i \in {\cal N}} {{{\left( {{R_i} - {D_i}} \right)}^2}} $ in ${{\cal P}_0}$ unification. Numerical verification results exceeds 1000, where per user traffic demand is randomly distributed with each simulation. To evaluate the proposed algorithm effect in a realistic level, the longitude range and latitude range of LEO constellation campaign are [$100^{\circ} \it{E}$ , $105^{\circ} \it{E}$ ] and [$-1.5^{\circ} \it{S}$ , $1.5^{\circ} \it{N}$ ], respectively. Actually, the six LEO satellites were randomly separated in three orbits over the region, where various geographic environments, i.e., oceans, rivers and plains, are included\cite{Lin2023}. Hence the complex geography caused by uneven ground station deployment and user traffic demand, which is considered more typical for R-NOMA-BF system. The relay satellite is positioned near the equator at $103^{\circ} \it{E}$. The simulated users are configured as a small number of satellite-enabled mobile terminals mounted on cars, boats, and drones \cite{Wang2023air}. Satellites related parameters and cooperative connections are referred to 3GPP TR 38.811 and TR 38.821 \cite{3GPP811,3GPP821}. The parameter definitions are summarized in TABLE I unless stated otherwise.
\begin{table}[!t]
\centering
\caption*{TABLE I: Simulation parameters}
\tabcolsep5pt
\renewcommand\arraystretch{1.2} 
\begin{tabular}{|l|l|}
\hline
\makecell[c]{Parameter}  &  \makecell[c]{Value} \\
\hline
\makecell[c]{Frequency, ${f_c}$}  & \makecell[c]{11.7 GHz}   \\
\hline
\makecell[c]{Bandwidth, ${B_{m{t_k}}}$}  &  \makecell[c]{500 MHz} \\
\hline
\makecell[c]{Satellite covers the ground longitude range}  &  \makecell[c]{[$100^{\circ} \it{E}$ , $105^{\circ} \it{E}$ ]} \\
\hline
\makecell[c]{Satellite covers the ground Latitude range}  &  \makecell[c]{[$-1.5^{\circ} \it{S}$ , $1.5^{\circ} \it{N}$ ]} \\
\hline
\makecell[c]{LEO satellites travel speed}   & \makecell[c]{7.9 km/s}  \\
\hline
\makecell[c]{LEO satellites altitude}    &  \makecell[c]{1200 km} \\
\hline
\makecell[c]{Relay satellite location}   & \makecell[c]{$103^{\circ} \it{E}$} \\
\hline
\makecell[c]{Relay satellite altitude}    &  \makecell[c]{36000 km} \\
\hline
\makecell[c]{Number of LEO satellites, $M$ }   &  \makecell[c]{6}  \\
\hline
\makecell[c]{User receive antenna gain, $G_{r}$ } & \makecell[c]{35.7 dBi } \\
\hline
\makecell[c]{LEO satellite maximum transmit \\ antenna gain, $G_{\max }$ }  &  \makecell[c]{64.9 dBi}  \\
\hline
\makecell[c]{Number of beamforming antennas, $L$}  & \makecell[c]{ 8}  \\
\hline
\makecell[c]{Number of cells, $K$ } &  \makecell[c]{64}  \\
\hline
\makecell[c]{Number of users per cell, $N$}  & \makecell[c]{ 4 } \\
\hline
\makecell[c]{Noise power, $\sigma^2$} & \makecell[c]{-136 dBW}  \\
\hline
\makecell[c]{Power budget per LEO satellite, $P_{s}$} & \makecell[c]{25 dBW} \\
\hline
\makecell[c]{Antennas array distance} & \makecell[c]{0.5 m} \\
\hline
\makecell[c]{Minimum capacity, $R_i^{\min }$} & \makecell[c]{5 Mbps} \\
\hline
\makecell[c]{Traffic demand, $D_i$} & \makecell[c]{300 Mbps to 1300 Mbps} \\
\hline
\end{tabular}
\label{parameter}
\end{table}

We verify the excellence of proposed algorithms by selecting several different comparison benchmarks as follows:
\begin{enumerate}
\item A-mNOMA-BF: Based on the scheme presented in Sections IV and V, A-mNOMA-BF algorithm is reintegrated. More specifically, the modified ant colony algorithm is combined with the monotone planning algorithm of NOMA. BF vector utilizes the SVD algorithm. The optimized variables are iterated over each other until convergence.
\item D-eNOMA-BF: Similar to A-mNOMA-BF algorithm, D-eNOMA-BF algorithm was redesigned as a new benchmark. The optimization results of BF vector are obtained first. Relay satellite measures the Doppler shift with the LEO satellites and passes the threshold constraints. The power obtained by exponential planning and obtained LEO satellite-cell matching factor are jointly optimized. The three variables are iterated to produce the final users' satisfaction.
\item Orthogonal multiple access beam forming (OMA-BF): Since each beam includes only a user, OMA-BF can be defined as a single-beam problem. The BF is still optimized according to the algorithm designed in (9).
\item  Relay satellite non-orthogonal multiple access 2 color (R-NOMA-2c): The subset of channels using the same polarization is called a color. The scheme refers to the consideration of orthogonal polarizations, i.e., right-handed or left-handed circular polarizations. The polarization multiplexing is considered in the existing LEO satellite-cell and NOMA algorithms without BF vector, which includes D-eNOMA-2c, A-eNOMA-2c, D-mNOMA-2c and A-mNOMA-2c. The power is distributed within the same polarization type \cite{Couble2018}.
\item Relay satellite non-orthogonal multiple access 4 color (R-NOMA-4c): Adding frequency multiplexing based on the R-NOMA-2c, associated algorithms contains four sub-bands consisting of different polarizations and frequencies, that is, D-eNOMA-4c, A-eNOMA-4c, D-mNOMA-4c and A-mNOMA-4c. Obviously, half of the bandwidth exists reduced in these cases.
\item Relay satellite non-orthogonal multiple access single beam (R-NOMA-S): A single beam gets to be designed into system, where all the antennas form a spot beam covering all users. Similarly this comparison baseline involves D-eNOMA-S, A-eNOMA-S, D-mNOMA-S and A-mNOMA-S \cite{Li2004}.
\end{enumerate}
For testing multiple objective functions on optimization results, we applied to discuss on the basis of four existing algorithms the following two schemes:
\begin{itemize}
\item Scheme 1: The objective function retains the second-order difference function from ${{\cal P}_0}$.
\item Scheme 2: The ratio between maximum reachable capacity and user demand is formulated in R-NOMA-BF system, while limiting achievable capacity  not exceed  traffic requirements, i.e., $\max \!\!\!\! \sum\limits_{k \in {\cal K},i \in {\cal N}} \!\! {\frac{{{R_i}}}{{{D_i}}}} ,{\rm{s}}{\rm{.t}}{\rm{.}} \ {R_i} \le {D_i}$ \cite{Lin2023}.
\end{itemize}

\subsection{Performance Analysis}
\begin{figure}[t!]
    \begin{center}
        \includegraphics[width=3.8in,  height=2.8in]{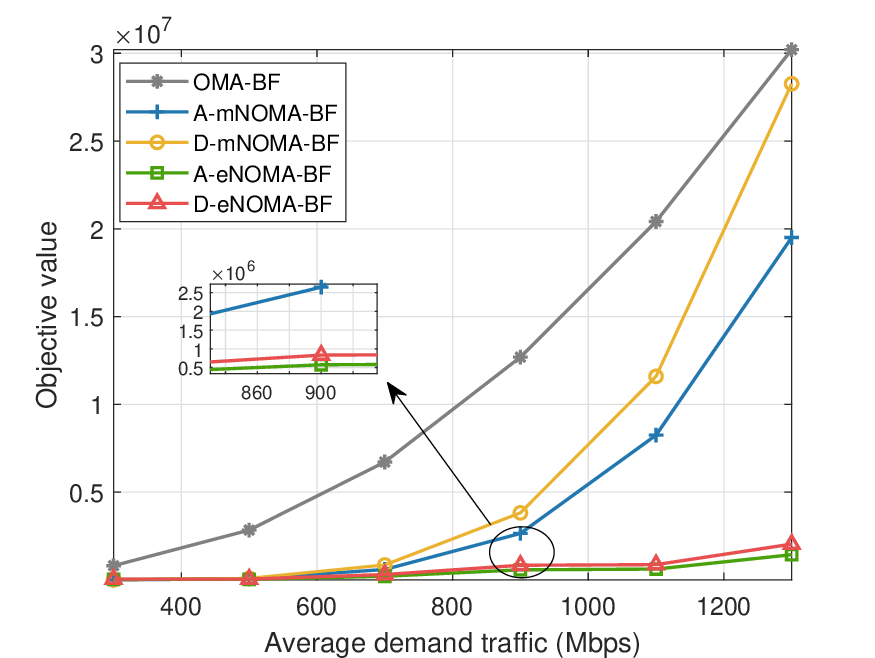}
        \caption*{Fig. 2: Objective gap versus average demand traffic with various comparative algorithms in R-NOMA-BF system.}
        \label{The_SOP_EE_diff_IR}
    \end{center}
\end{figure}

Fig. 2 plots the objective gap versus average demand traffic with various comparative algorithms for OMA-BF, A-mNOMA-BF, D-mNOMA-BF, A-eNOMA-BF and D-eNOMA-BF. All four different algorithms have excellent control within the user demand of 500 Mbps. However, these algorithms expose significant disparities beyond 600 Mbps. As can be observed that eNOMA provides a more robust fit compared to mNOMA. In the case of demand growth from 700 to 1300 Mbps, A-eNOMA-BF and D-eNOMA-BF algorithms only dropped by $14.12\% $ and $14.90\% $, respectively. The improved ant colony based algorithm better achieves global planning based on the optimization results calculated from a large number of pheromone. However, the Doppler shift algorithm only focuses on the optimal matching results for each LEO satellite, which leads to some differences between two types of algorithms in fitting ability. Since two algorithms just illustrated are excellent, the ant colony algorithm still improves matching performance, which is more valuable. Compared with that, A-mNOMA-BF and D-mNOMA-BF algorithms have a cliff-like decline in fitting gaps, but the revised ant colony algorithm has an ameliorative role. This phenomenon indicates that the approximation of ${{{\cal P}'_1}}$ in the solution process leads to loss of substantial accessible capacity.
\begin{figure}[t!]
    \begin{center}
        \includegraphics[width=3.8in,  height=2.8in]{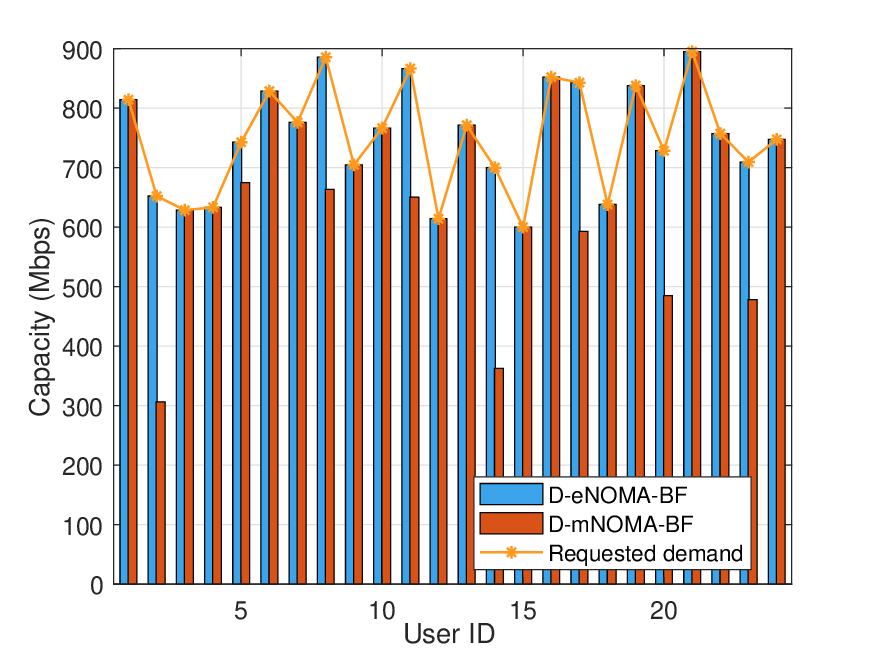}
        \caption*{Fig. 3: User ID versus achievable capacity for Doppler frequency shift LEO satellite-cell assignment strategy.}
        \label{The_SOP_EE_diff_IR}
    \end{center}
\end{figure}

Fig. 3 plots the user ID versus achievable capacity with D-eNOMA-BF and D-mNOMA-BF algorithms. The blue and green bars represent the achievable capacity of the D-eNOMA-BF and D-mNOMA-BF algorithms, respectively. The orange curve is the users' traffic request. One observation is that D-eNOMA-BF algorithm essentially allows the two types of results to overlap. In contrast, nearly $30\%$ users  differed significantly in D-mNOMA-BF algorithm, where user 14 has the largest disparity of 337.37 Mbps. This phenomenon shows that D-eNOMA-BF algorithm still fully achieves optimal user satisfaction at nearly 700 Mbps user demand. We conclude that the exponential cone programming provides better user satisfaction for independent individual users. In D-mNOMA-BF algorithm, the approximation accuracy of the optimization function becomes lower for too large an auxiliary variable $\partial$ about $p_{\max ,i}$ based on (17a). The objective function is actively split and approximated into two logarithmic functions according to (13) and (14), which can narrow the desirable range of $p_i$. As a result, D-mNOMA-BF algorithm appears some traffic misfit for high traffic request of users.
\begin{figure}[h!]
    \begin{center}
        \includegraphics[width=3.8in,  height=2.8in]{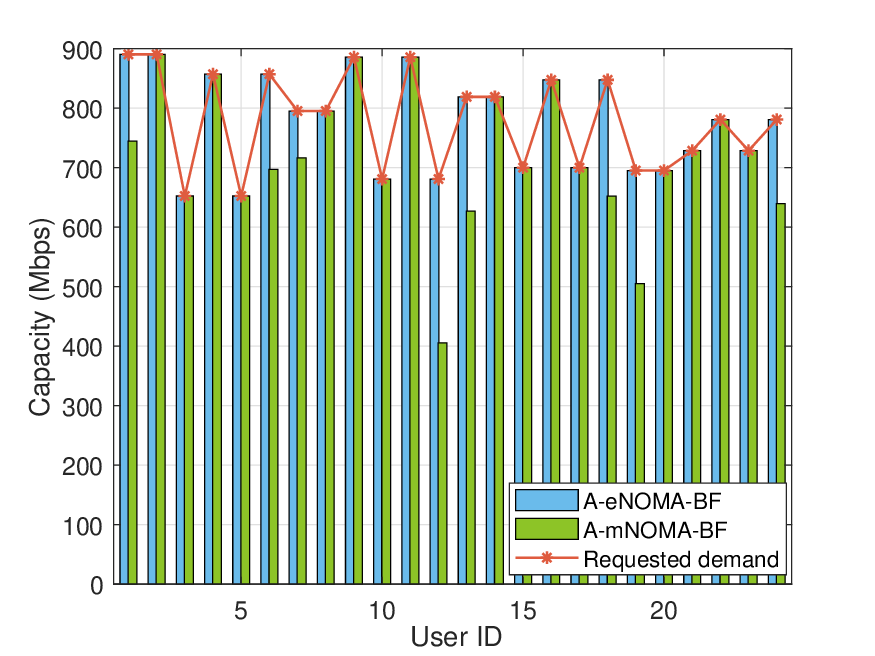}
        \caption*{Fig. 4: User ID versus achievable capacity for modified ant colony algorithm.}
        \label{The_SOP_EE_diff_IR}
    \end{center}
\end{figure}

As a further advance, Fig. 4  plots the user ID versus achievable capacity for A-eNOMA-BF and A-mNOMA-BF algorithms. The blue bar and red bar indicate the achievable capacity of the users in A-eNOMA-BF and A-mNOMA-BF algorithms, respectively. The yellow curve represents the request traffic of users. It is worth mentioning that about $30\%$ users have not achieved their satisfaction in A-mNOMA-BF algorithm. The maximum distance between user 12 and its traffic demand is 275.51 Mbps. Similar with the principle of Doppler shift based algorithm, the performance of A-mNOMA-BF algorithm is still weaker than that of A-eNOMA-BF. It is worth noting that the improved ant colony algorithm for LEO satellite matching cell indexing method is different from Doppler shift based algorithm,  which causes user request traffic differentiation. The average gap of users for A-mNOMA-BF and D-mNOMA-BF algorithms are 208.12 Mbps and 301.14 Mbps, respectively. The A-mNOMA-BF algorithm outperforms D-mNOMA-BF algorithm by 0.45 dB. Over the critical 600 to 900 Mbps range, eNOMA-based algorithms are largely fitted for capacity and request traffic.
\begin{figure}[h!]
    \begin{center}
      \subfigure{
    \begin{minipage}[b]{0.45\textwidth}
        \includegraphics[width=3.8in,  height=2.8in]{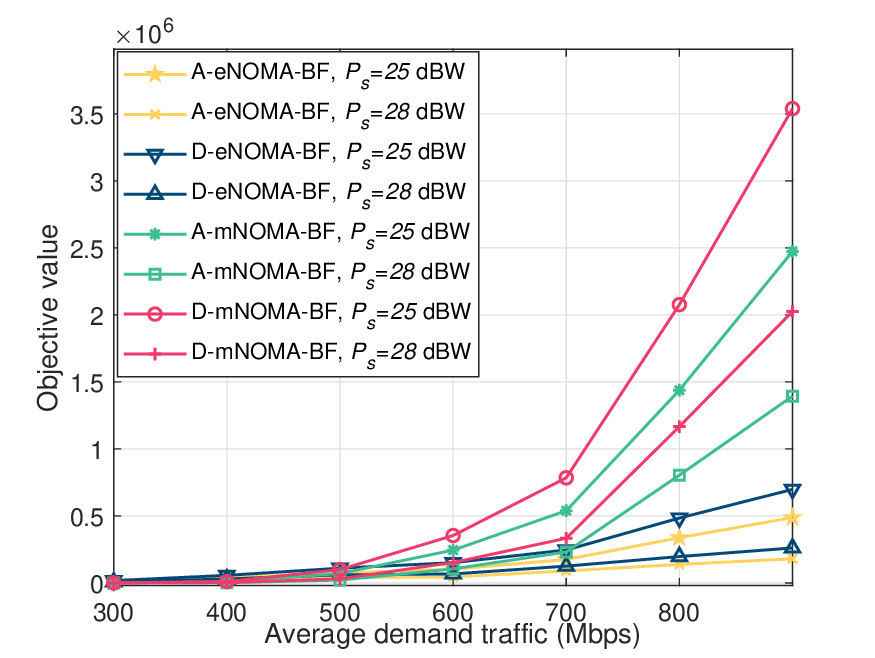}
       \caption*{(a) The satisfaction of users versus average demand traffic with different power in R-NOMA-BF system.}
        \end{minipage}
        }
        \subfigure{
        \begin{minipage}[b]{0.45\textwidth}
          \includegraphics[width=3.8in,  height=2.8in]{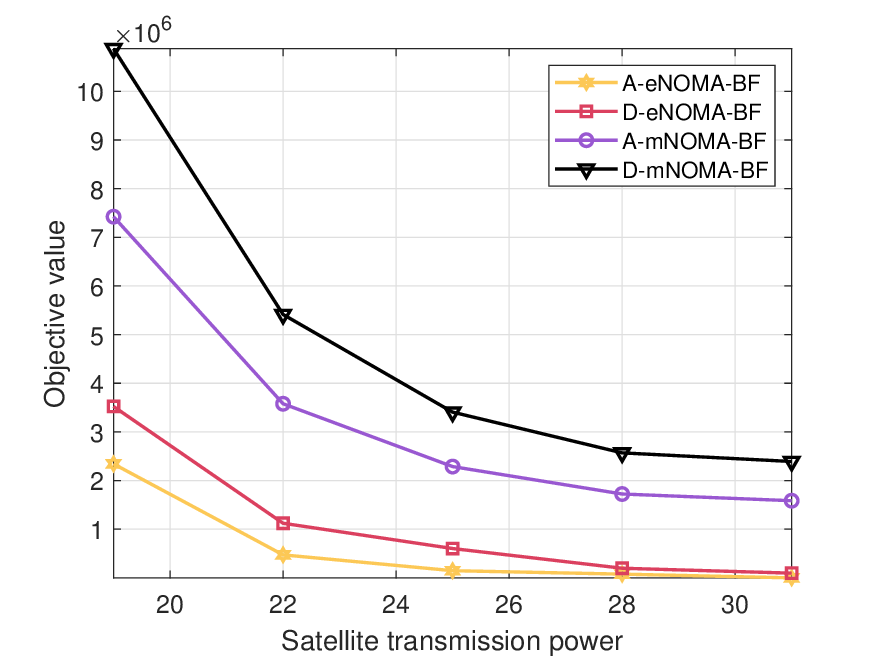}
          \caption*{(b) The satisfaction of users versus satellite transmission power with 900 Mbps of average user demand.}
        \end{minipage}
        }
         \caption*{Fig. 5: Simulation results for user satisfaction and satellite transmission power.}
    \end{center}
\end{figure}

Fig. 5(a) plots the traffic gap related to user demand with multiple power in R-NOMA-BF system,  where ${P_s} = 25 \ {\rm{ dBW}}$ or $28 \ {\rm{ dBW}}$. We can observe that single satellite power more higher the overall solution of user satisfaction are enhanced under power constraints. With the increasing of ${P_s}$ increasing, A-eNOMA-BF and D-eNOMA-BF algorithms are simultaneously  improved by 3 dB in matching gap at 900 Mbps. The fitting ability of both A-mNOMA-BF and D-mNOMA-BF algorithms are also boosted by 2.67 dB with power. This suggests that not only the power allocation per NOMA users is essential, but also optimization in the directional power to BF vector affects users capacity. Therefore, NOMA and BF vector power assignments make sense to be studied on the basis of payload limitations for LEO satellite cooperative communication. As a further advance, Fig. 5(b) plots the curve of user satisfaction with respect to the satellite transmission power ${P_s}$, where the average user demand is 900 Mbps. As can be observed that user satisfaction is not proportional to the increase in transmit power. The formulated objective scheme concentrates on satisfying the accomplishable user demand for lower ${P_s}$, which makes the relative majority of users to be assigned less power allocation. As a result, users with higher satisfaction are subjected to fewer interferences, where boosting the transmit power leads to larger gains. With the higher transmit power, user demand is basically satisfied, and this further increase in ${P_s}$ will not enhance the performance much.
\begin{figure}[h!]
    \begin{center}
        \includegraphics[width=3.8in,  height=2.8in]{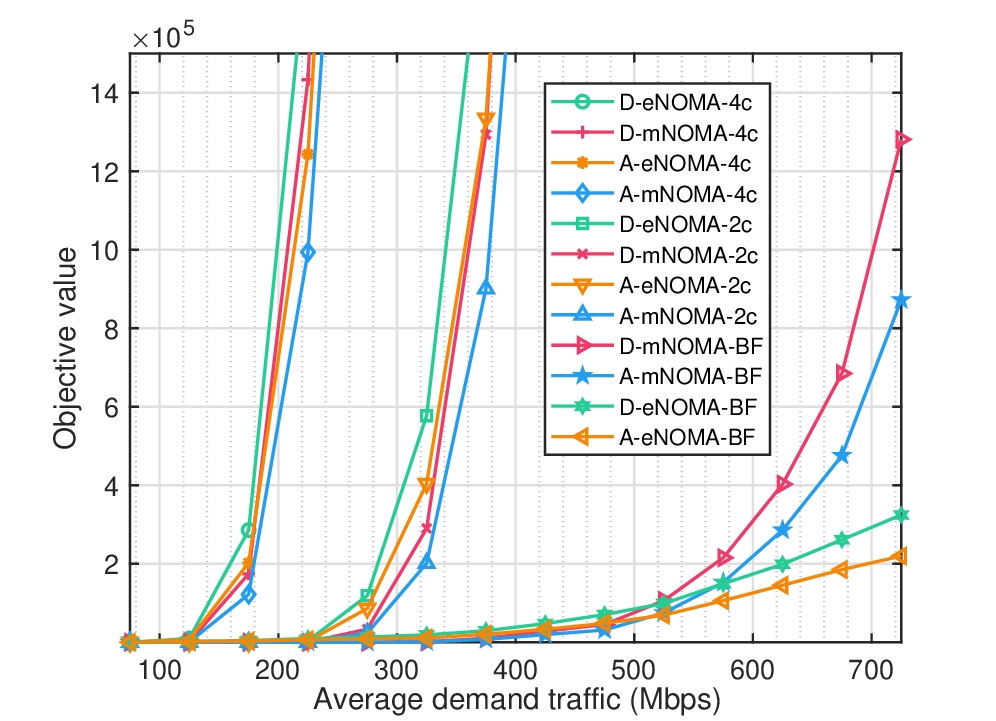}
        \caption*{Fig. 6: Comparison of user satisfaction gap versus average demand traffic for multiple polarization modes.}
        \label{The_SOP_EE_diff_IR}
    \end{center}
\end{figure}

To evaluate the impact of BF vector and antenna polarization on R-NOMA-BF system performance, Fig. 6 compares the gap among multiple polarization modes of user satisfaction versus average traffic demand. BF weights and synthesizes the signals received from multiple antenna array elements in all directions. BF is capable to focus on a specific direction and mitigate interference for surrounding users. However, polarization multiplexing is the forward and backward of the phase angles between the electric and magnetic field components to define left- and right-handed polarization.  The users of different polarization types have hardly interference with each other. For the sake on convenience, it is emphasized that main algorithms refer to 1-color multiplexing, i.e., users occupy the entire frequency band. As can be observed that the optimal user requested traffic for 4-color multiplexing lie only around 130 Mbps, while 2-color multiplexing approach 230 Mbps at the most appropriate user satisfaction. It illustrates that 4-color multiplexing sacrifices bandwidth to save inter-user interference and hardly has a nice answer. The correlation algorithms with BF vector do not waste bandwidth despite controlling the interference, which highlights its performance. Furthermore, the optimization without BF vector causes little division across the four algorithms based on polarization modes. Such phenomenon shows that the power derived from mNOMA and eNOMA approximation no longer has negative consequences on ${p_i}$ of BF vector by (9). In Table II, we further summarize the proposed schemes for jumping out the optimal satisfaction bound capacity. Due to 4-color multiplexing sharing the frequency resources, these algorithms reduce achievable rates. Compared to BF-based algorithms, 2-color multiplexing only isolates users' interference of different polarization types, which still adversely affects traffic fitting. However, the BF-based algorithms improve users' satisfaction by isolating interference in all directions. The total actual capacity of users represents the limit of traffic without spilling over. It is concluded that the BF-based algorithms effectively enhance the upper bound of users' satisfaction.
\begin{table}[!h]
\centering
\caption*{TABLE II: Total actual users capacity and unsatisfied users traffic gap of algorithms with various polarization modes}
\tabcolsep5pt
\renewcommand\arraystretch{1.2} 
\begin{tabular}{|l|l|l|}
\hline
\makecell[c]{Algorithm}  &  {\makecell{Total actual users \\  capacity (Mbps)}}  & {\makecell{Total unsatisfied users \\   traffic gap (Mbps)}}  \\
\hline
\makecell[c]{D-eNOMA-4c} &   \makecell[c]{50334} & \makecell[c]{16982} \\
\hline
\makecell[c]{A-eNOMA-4c}  &  \makecell[c]{52981} & \makecell[c]{14221} \\
\hline
\makecell[c]{D-mNOMA-4c} & \makecell[c]{54018} & \makecell[c]{13196} \\
\hline
\makecell[c]{A-mNOMA-4c}  & \makecell[c]{56167} & \makecell[c]{11049} \\
\hline
\makecell[c]{D-eNOMA-2c}   & \makecell[c]{94711} & \makecell[c]{10984}  \\
\hline
\makecell[c]{A-eNOMA-2c}    &  \makecell[c]{96330} & \makecell[c]{9279} \\
\hline
\makecell[c]{D-mNOMA-2c}   & \makecell[c]{99837} & \makecell[c]{5771} \\
\hline
\makecell[c]{A-mNOMA-2c}    &  \makecell[c]{100775} & \makecell[c]{4832} \\
\hline
\makecell[c]{D-mNOMA-BF}    &  \makecell[c]{174011} & \makecell[c]{8390}  \\
\hline
\makecell[c]{A-mNOMA-BF}  & \makecell[c]{175472} & \makecell[c]{6928}  \\
\hline
\makecell[c]{D-eNOMA-BF}  &  \makecell[c]{175629} & \makecell[c]{6771}  \\
\hline
\makecell[c]{A-eNOMA-BF}  & \makecell[c]{176813} & \makecell[c]{5587} \\
\hline
\end{tabular}
\label{parameter}
\end{table}

\begin{figure}[h!]
    \begin{center}
        \includegraphics[width=3.8in,  height=2.8in]{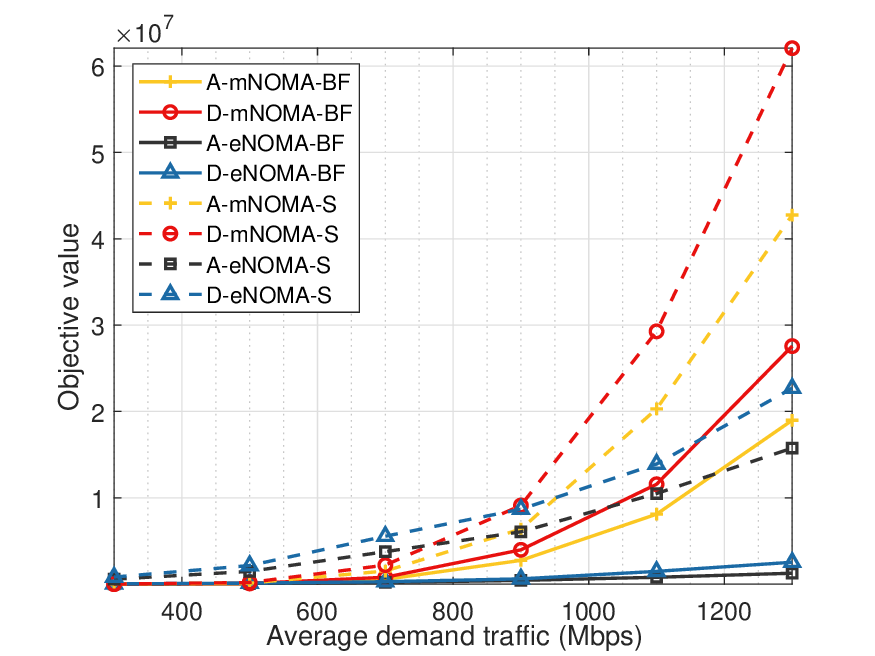}
        \caption*{Fig. 7: Satisfaction objective value versus user demand traffic with single beam benchmarks.}
        \label{The_SOP_EE_diff_IR}
    \end{center}
\end{figure}

Fig. 7 plots the users' satisfaction of proposed algorithms for comparing the conventional single-beam multi-antenna scenarios. The firm and dotted lines indicate four algorithms already available above and the single-beam comparison baselines, respectively. It shows that eNOMA-S has inferior performance than mNOMA-S, where the former needs to be close at 300 Mbps before completing users demand, whereas the latter takes 500 Mbps. This is because that the isolated interference based on different directional power is weakened by BF vector optimization in multi-antenna single beam. The strong power distribution characteristics of eNOMA cannot be reflected as a result. Another observation is that mNOMA grows more steeply after leaving the optimal satisfaction. This is due to the fact that auxiliary variables about ${p_i}$  are no longer within the constraints by (20). Due to the limited space resources available for satellite communications, the scenarios of a single beam serving multiple users are often seen in the same direction. R-NOMA-S system is a special case of R-NOMA-BF system. R-NOMA-S systems have more focused radiant energy, while R-NOMA-BF systems are more concerned with suppressing interference to improve performance. This means that the inter-beam interference neglect of BF vector in exponential planning has a greater impact on eNOMA algorithms. Therefore, mNOMA is a better choice in multi-antenna single beam scenarios.
\begin{figure}[t!]
    \begin{center}
        \includegraphics[width=3.8in,  height=2.8in]{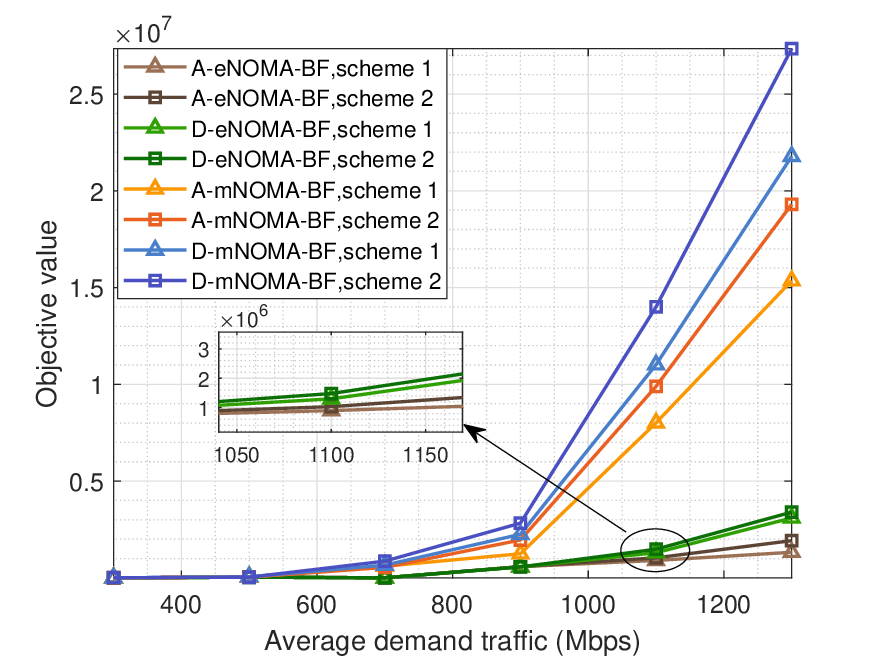}
        \caption*{Fig. 8: The performance of gap versus user demand traffic for different objective function schemes.}
        \label{The_SOP_EE_diff_IR}
    \end{center}
\end{figure}

Fig. 8 plots the performance of gap versus user demand traffic for different objective function schemes in R-NOMA-BF system. We can observe that the schemes of transferring  objective problem from  difference function to ratio function make the gaps larger. The constraint ${R_i} \le {D_i}$ of scheme 2 has a significant influence on the natural fit compared to 1 for mNOMA. More specifically, the traffic gap of D-mNOMA-BF algorithm changes upward by 1.10 dB from scheme 1 to 2, and A-mNOMA-BF rises at 1.02 dB. The other phenomenon is that both A-eNOMA-BF and D-eNOMA-BF algorithms do not differ much, yet they respectively improved in 0.75 dB and 0.66 dB  for both schemes as users demand increased up to 1100 Mbps. Accordingly, the second-order function of discrete difference better suits the algorithms proposed above. Scheme 2 is restricted to ${R_i} \le {D_i}$, indicating that users' satisfaction is weakened in the case of slight lost traffic. The proposed algorithms show good adaptive and generalization capabilities among different metrics.
\begin{figure}[h!]
    \begin{center}
        \includegraphics[width=3.8in,  height=2.8in]{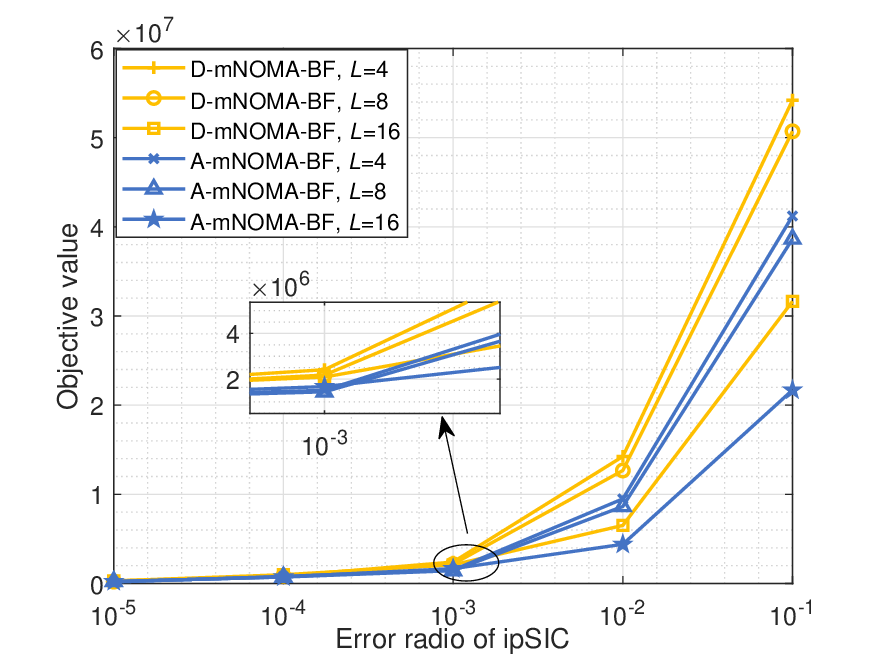}
        \caption*{Fig. 9: The objective traffic gap versus error radio of ipSIC with multiple antenna numbers.}
        \label{The_SOP_EE_diff_IR}
    \end{center}
\end{figure}
\begin{figure}[h!]
    \begin{center}
        \includegraphics[width=3.8in,  height=2.8in]{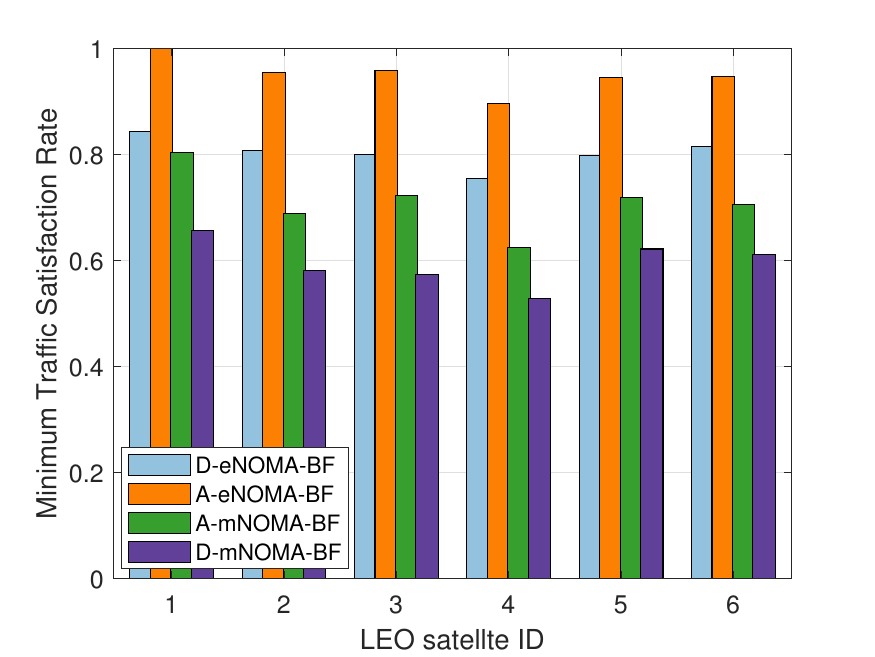}
        \caption*{Fig. 10: The minimum traffic satisfaction rate versus satellite ID with various comparative algorithms in R-NOMA-BF system.}
        \label{The_SOP_EE_diff_IR}
    \end{center}
\end{figure}

Fig. 9 plots objective traffic gap versus error radio of ipSIC with multiple antenna numbers, where $0 \le {\kappa _i} \le 1$ is introduced to denote ipSIC factor. The user interference of the first decoding is not completely eliminated in the cell. Hence the interference of users in R-NOMA-BF system is rewritten as \cite{HOUtianwei}
\begin{align}\label{29}
\sum\limits_{\mathop {j \in {\cal N}\backslash \{ i\} }\limits_{j < i} } \!\!\! {{{\left| {{\bf{h}}_i^H{{\bf{w}}_j}} \right|}^2}{p_j}}  + \sum\limits_{\mathop {j \in {\cal N}\backslash \{ i\} }\limits_{j > i} } \!\!\! {{{\left| {{\bf{h}}_i^H{{\bf{w}}_j}} \right|}^2}{p_j}{\kappa _i}}  .\tag{29}\nonumber
\end{align}
Since the exponential cone programming cannot iterate power ${p_i}$ with pSIC through (26), the yellow and blue curves are plotted for D-mNOMA-BF and A-mNOMA-BF algorithms from antenna number 4, 8 to 16, respectively. As observed can be that when ${{\kappa _i}}$ exceeds $10^{-3}$, D-mNOMA-BF and A-mNOMA-BF algorithms with $L=16$ become a better option. Elevated ${{\kappa _i}}$ means that target users receive more interference from decoding users first. Besides, the lower number of antennas is unable to make a dramatic change in the performance of A-mNOMA-BF algorithm. This phenomenon shows that antenna number affects BF vector optimization enabling more accurate interference isolation, where energy in different directions is concentrated according to (9). As the array response vectors get more accurate, user capacities are raised. The increased number of antennas takes into account the phased array antenna size and layout changes on the satellite side, where the size of the flat plate, the number and  arrangement of array elements  are included. In addition, intelligent satellite processors are regarded as an important method to enhance data processing capability.
\begin{figure}[h!]
    \begin{center}
        \includegraphics[width=3.8in,  height=2.8in]{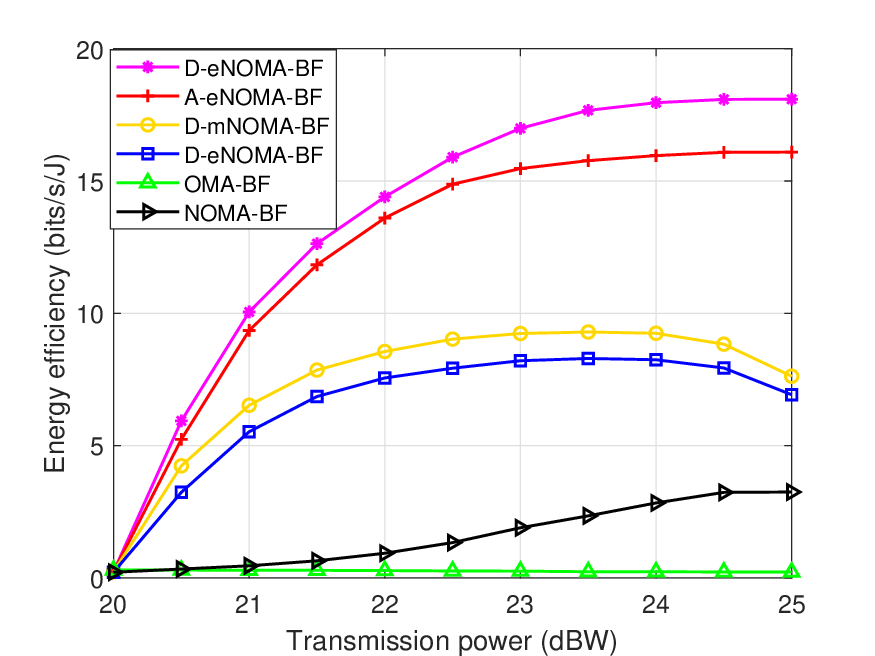}
        \caption*{\textcolor[rgb]{0.00,0.00,1.00}{Fig. 11: Energy efficiency versus satellite transmission power with different algorithms.}}
        \label{The_SOP_EE_diff_IR}
    \end{center}
\end{figure}

Fig. 10 plots the bar graph of the minimum traffic satisfaction rate versus LEO satellite ID in R-NOMA-BF system. The minimum traffic satisfaction rate is defined as $\mathop {\max }\limits_{\cal K} \left\{ {\min \left\{ {\frac{{{R_i}}}{{{D_i}}},1} \right\}} \right\}$ for each LEO satellite, which implies the worst capacity-demand mismatch between the LEO satellite and users. It can be observed that A-eNOMA-BF algorithm outperforms the other benchmarks with high traffic satisfaction for each satellite. Moreover, in A-eNOMA-BF algorithm, the minimum traffic satisfaction rate for LEO satellite 1 is one, i.e., all users in satellite 1 are satisfied. A lower minimum traffic satisfaction rate indicates a smaller percentage of satisfied users for the satellite service. In addition, the modified ant colony algorithms all outperform the performance of doppler shift based algorithms. A conclusion can be drawn that modified ant colony algorithm is more effective in global optimization.

\textcolor[rgb]{0.00,0.00,1.00}{Fig. 11 plots the energy efficiency versus satellite transmission power with different algorithms in R-NOMA-BF system. The energy efficiency is defined as the ratio of the information transmission rate to the transmit power, i.e.,$\sum\limits_{m \in {\cal M}} {\frac{{\sum\limits_{i \in {\cal N}} {\log } \left( {1 + {\gamma _i}} \right)}}{{P_m^{tot}}}} $.  ${{P_m^{tot}}}$ represents the sum of power allocated to  users by each satellite. As can be observed that with increasing satellite transmission power, D-eNOMA-BF and A-eNOMA-BF algorithms approach a maximum of energy efficiency at the power of 25 dBW. This indicates that the energy efficiency cannot always keep growing with power due to the indirect effects of user traffic requests and interference. In addition, in contrast to OMA and LEO satellite communication without relay satellite assistance algorithms, the proposed algorithms have significantly higher performance in terms of energy efficient. The relay satellite can reduce the energy consumption of LEO satellite nodes and ground stations, while the NOMA-based algorithms can effectively avoid the waste of power resource.}
\section{Conclusion}
In this paper, the performance of R-NOMA-BF system has been studied, where the relay satellite assists LEO constellation with multiple antennas covering multiple users. The optimization problems for users' satisfaction and constraint on power, LEO satellite-cell matching factor as well as BF vector constraints have been formulated. According to above conclusions, the procedure and complexity of D-mNOMA-BF and A-eNOMA-BF algorithms were given. Numerical results have shown that the demonstrated algorithms have sufficient advantages over the traditional comparison benchmarks. We further validated that antenna number and single-satellite power have improved system performance. The impact of interference generated by ipSIC on users satisfaction has been investigated.  From the perspective of practical applicability, R-NOMA-BF system is capable of satisfying the need for more satellite full coverage, where the ground users can be sea or large unoccupied area terminals.  \textcolor[rgb]{0.00,0.00,1.00}{The setup of perfect CSI and more numbers of LEO satellites and users brings about high performance for the proposed algorithm, hence our future work will consider the impact of imperfect CSI and seek design methods for full coverage of the ground surface.} The efficient processing tools and green communication can further accomplish meaningful contributions into R-NOMA-BF system, which is a very practical future investigation topic. 

\bibliographystyle{IEEEtran}
\bibliography{mybib}

\end{document}